\documentstyle[epsf,12pt]{article}
\newcommand{\al}{\alpha'}
\newcommand{\de}{\partial}
\newcommand{\be}{\begin{equation}}
\newcommand{\ba}{\begin{eqnarray}}
\newcommand{\ea}{\end{eqnarray}}
\newcommand{\ee}{\end{equation}}
\newcommand{\db}{\bar{\partial}}

\newcommand{\ca}{\mathcal}
\newcommand{\lr}{\leftrightarrow}
\newcommand{\f}{\frac}
\newcommand{\s}{\sqrt}
\newcommand{\vp}{\varphi}

\newcommand{\tvp}{\tilde{\varphi}}
\newcommand{\ti}{\tilde}
\newcommand{\ap}{\alpha}

\newcommand{\mb}{\mathbf}
\newcommand{\ddd}{\cdot\cdot\cdot}
\newcommand{\no}{\nonumber \\}

\newcommand{\lb}{\rangle}

\renewcommand{\ss}{\scriptscriptstyle}

\begin{document}
\begin{titlepage}
\thispagestyle{empty}
\begin{flushright}
hep-th/0110200 \\
UT-972 \\
October, 2001 \\
\end{flushright}

\bigskip
\bigskip

\begin{center}
\noindent{\Large \textbf{D-branes in Melvin Background}}\\
\bigskip
\bigskip
\noindent{
          Tadashi Takayanagi\footnote{
                 E-mail: takayana@hep-th.phys.s.u-tokyo.ac.jp}
                 and Tadaoki Uesugi\footnote{
                 E-mail: uesugi@hep-th.phys.s.u-tokyo.ac.jp} }\\

\bigskip
\it{Department of Physics, Faculty of Science \\ University of Tokyo \\
\medskip
Tokyo 113-0033, Japan}

\end{center}
\begin{abstract}
In this paper we discuss D-branes in the Melvin background and its 
supersymmetric generalizations. In particular we determine the D-brane spectra 
in these backgrounds by constructing their boundary states explicitly, where 
some of the D-branes are supersymmetric. The results sensitively depend on 
whether the value of magnetic flux in the Melvin background 
is rational or irrational. For the rational case the D-branes are regarded as 
the generalizations of fractional D-branes 
in abelian orbifolds ${\bf C^n/Z_N}$ of type II or type 0 
string theory. For the irrational case we found a very limited spectrum. 
Since the background includes the nontrivial H-flux, the D-branes will 
provide interesting examples from the viewpoint of the noncommutative geometry.
\end{abstract}
\end{titlepage}

\newpage

\section{Introduction}
\setcounter{equation}{0}
\hspace{5mm}

One of the important problems in string theory is to understand 
its vacuum structure completely. Some recent discoveries have revealed the
connections between the several vacua whose definitions were originally
thought to be different from each other. The first great advance 
came with the discovery of the string duality \cite{Wi0}.
The five string `theories' in flat space turned out to be 
not different theories but different vacua in a single theory, 
that is to say, M-theory. The second advance has been made from the 
investigation of 
the open string tachyon condensation on unstable and thus 
non-supersymmetric D-brane systems (for a review see \cite{Sen994,Oh}). 
The key idea of understanding this phenomenon is the Sen's conjecture, 
which says that the unstable vacuum with D-branes decays 
to stable lower dimensional D-branes or the closed string vacuum.
{}From this we can expect that the original vacuum with the unstable 
D-brane system belongs to the {\sl spontaneous} supersymmetry broken 
phase. Indeed the nonlinear supersymmetries can be realized on 
such systems \cite{Yo,TeUe}.  

{}From this we expect that the above observation about open 
string unstable systems may be applied to closed string 
unstable vacua such as type 0 string etc. (for the type 0 theory see
\cite{Po}).
If we naively translate the above statement into those examples,
such closed string `theories' are not different theories from 
type II but only different vacua. One attempt in this direction will be
the conjecture that type 0A string is equivalent to ${\bf S}^1$ 
compactification of M-theory with the anti-periodic boundary condition for
all fermions \cite{BG}. Moreover we may speculate that 
the type 0 string is a spontaneously broken phase of the original 
thirty two supersymmetries in type II, where the gravitino may obtain 
the infinite mass and thus disappear. However, in reality it seems 
difficult to 
explain it completely within the framework of classical 
supergravity. For example,
one should explain the presence of tachyon in type 0 string and the 
doubling of RR-fields. 
These intriguing questions about unstable closed string vacua will also
be closely related to the closed string tachyon condensation, which
remains to be well understood (for latest discussions see e.g 
\cite{AS,RM,AdPoSi,Da,Su2,TaUe,RuTs}).

Recently the Kaluza-Klein Melvin backgrounds \cite{Me,GiWi,GiMa,Dowker} have 
been intensively studied both in M-theory (or superstring with RR flux)
\cite{CG,Sa2,RM,GS,CoHeCo,Em,Sa1,BrSa,RM2,Ur, FiFa} and in 
NS-NS superstring model
\cite{SM,SM2,CG,Su2,TaUe,RuTs}. Both backgrounds will 
provide useful materials for the studies of unstable closed string vacua
since these are known to connect type 0 string with 
type II string in the flat space \cite{CG}.
In the former context the Melvin-like 
solutions are called flux-branes (Fp-brane) \cite{Sa2,RM,GS,CoHeCo}. 
On the other hand, the latter model gives an exactly solvable superstring 
model in spite of its non-trivial curved background with $H$-flux
\cite{SM,SM2,TaUe,RuTs}.
The simplest model \cite{SM,SM2} (`9-11' flip of F7-brane) includes three 
parameters (the radius $R$ and two magnetic fluxes $q$ and $\beta$) and 
thus does cover many vacua. In particular,
by tuning these parameters appropriately we can realize the type 0 or 
type II string and this shows explicitly that  
the Melvin background connects type 0 with type II. Furthermore it can be
shown that all of the abelian orbifolds ${\bf C}/{\bf Z}_N$ in type 0 or 
type II string are also included as special limits \cite{TaUe}. 
We can also construct supersymmetric Melvin backgrounds \cite{TaUe,RuTs} 
by considering the higher 
dimensional generalizations  (`9-11' flip of F5, F3 and F1-brane). These
also include supersymmetric orbifolds such as ALE orbifolds
${\bf C}^2/{\bf Z}_N$ (for ALE space see \cite{ALE}) 
as well as nonsupersymmetic orbifolds. 
These Melvin models describe a large region 
of string vacua both supersymmetric and nonsupersymmetic, interpolating
various orbifolds. Therefore these models will be suitable to explore
the structure of vacua in string theory and the decay of unstable vacua.

If we are given a non-trivial background like one of these, 
a D-brane generally provides a good probe to
investigate its stringy geometry. Thus in this paper we will
discuss various aspects of D-branes in the Melvin backgrounds.
Because the model is exactly solvable as we have mentioned, 
the boundary conformal field theories of D-branes are also 
treated exactly. In particular we can construct their boundary states
explicitly as we will see later. As a result we find that 
these D-branes depend 
on the parameters of the Melvin background in intriguing ways.
For example, if we assume the simplest Melvin model \cite{SM,SM2} and
the parameters $qR$ and $\f{\beta\al}{R}$ are rational numbers,
the D-branes which wrap the compactified circle
depend only on $qR$, while those which are point-like along the circles
depend only on $\frac{\beta\al}{R}$. 
This difference sometimes causes the strange Bose-Fermi degeneracy
in the open string spectrum on the D-branes in spite of the absence of
supersymmetry in the closed string sector. We will later discuss an 
interpretation of this degeneracy as a remnant of the spontaneously broken
supersymmetry. 

Even though these two kinds of D-branes are 
transformed into each other by T-duality, they look different 
geometrically. The former kind of the D-brane has the spiral 
world-volume wrapping several times on two
dimensional tori like coils. On the other hand, the latter 
kind of the D-brane is very similar to fractional D-branes 
\cite{DoMo,DiDoGo} in orbifold 
theories. Indeed as we have mentioned, if we take appropriate limits,
then the model becomes equivalent to orbifolds ${\bf C}/{\bf Z}_N$ in
type 0 and type II string. We can see that the D-branes in 
the Melvin background really become the fractional D-branes in the 
orbifold limit. Especially we find that
a D-brane in the Melvin model is divided into an electric fractional 
D-brane and
a magnetic fractional D-brane \cite{KlTs,BiCrRo,BeGa2,BG,KlNeSh} 
in the type 0 limit.
We also find an interesting phenomenon that the type of a
fractional-like D-brane is changed due to a kind of monodromy 
if it goes around the circle. 

Moreover,
D-brane spectra are very sensitive to whether the magnetic parameters
$qR$ and $\frac{\beta\al}{R}$
are rational or irrational. It was found that if we take the decompactified
limit in the irrational case, then the closed string sector approaches not 
the ordinary orbifold ${\bf C}/{\bf Z}_N$ but an unfamiliar 
kind of a `large $N$ limit of the orbifold' \cite{TaUe}. In this case 
all of the allowed D-branes 
are pinned at the fixed point of the 
Melvin background and cannot move around, while in the rational case there
exist movable D-branes similar to 
those which belong to the regular representations
in the orbifold theories. Indeed for the irrational parameters 
if we would try to move the D-branes away 
which wrap the compactified circle, it should wind infinitely many
times (`foliation') and thus it becomes singular. 

In this way the D-branes in Melvin backgrounds change their aspects
dramatically in accordance with the various values of magnetic parameters.

It will also be an interesting result that if we consider the 
higher dimensional supersymmetric 
Melvin backgrounds \cite{TaUe,RuTs}, then we can obtain BPS D-branes. 
In the orbifold limit these D-branes are identified with 
BPS fractional D-branes on ALE orbifolds ${\bf C}^2/{\bf Z}_N$.

Finally we would like to mention another interesting viewpoint on the 
Melvin backgrounds. They include non-trivial $H$-flux for non-zero $\beta$ 
and thus the world-volume theory on the D-branes in these backgrounds may 
be analyzed in terms of noncommutative geometry \cite{CoDoSc,SeWi,CoSc}. 
Related examples from this viewpoint may be the D-branes in 
group manifolds \cite{KaOk,AlReSc}, 
where the non-trivial $H$-flux is also present. 
Recently the D-brane charges of 
these \cite{AlSc} are explained by employing twisted K-theory 
\cite{Witten3} in \cite{FrSc,MaMoSe}.

The plan of this paper is as follows.  In section two we review 
the Melvin background and the quantization of its sigma model 
by the operator method. In section three we investigate the D-branes 
in the Melvin background by using the boundary state formalism.
We calculate the vacuum amplitude and check the open-closed
duality relation. We discuss the geometrical properties of them. We also
examine the relation to fractional D-branes in orbifold theories.
In section four we generalize the arguments in 
section three to those of the higher dimensional Melvin model. We see that
the D-branes in this background become 
supersymmetric for specific values of the parameters.
In section five we draw conclusions and discuss the future problems.
In appendix we summarize some known facts which are necessary for the 
analysis in this paper.

\section{Closed Strings in the Melvin Background}
\setcounter{equation}{0}
\hspace{5mm}
In this section we mainly review the exactly solvable model of type II
superstring studied in \cite{SM,SM2} as well as some recent related
developments \cite{CG, BG, TaUe}.
See also \cite{BM2,BM3,BM} for such a model in bosonic string theory.
We also present the detailed analysis of the world-sheet fermions 
in NS-R formalism by using superfields.

The target space of this model has the structure of Kaluza-Klein 
theory
and has the topology $\mbox{M}_3\times{\bf R}^{1,6}$. 
The three dimensional 
manifold $\mbox{M}_3$ is given by ${\bf S}^1$ fibration over ${\bf R}^2$. 
We write the coordinate of ${\bf R}^2$ and ${\bf S}^1$ by $\rho,\vp$ 
(polar coordinate) and
$y$ (with radius $R$).  
This non-trivial fibration is due to two Kaluza-Klein (K.K.)
 gauge fields $A_{\vp}$
and $B_{\vp}$ (see eq.(\ref{metric})) 
which originate from K.K. reduction of metric 
$G_{\vp y}$ and B-field $B_{\vp y}$, respectively. 
Thus this background can be viewed as a generalization \cite{GiWi, GiMa} 
of the original Melvin solution \cite{Me} to string theory, which is 
recently discussed
in the context of the flux brane \cite{Sa2, GS, RM, CoHeCo, Em, Sa1,
BrSa, Ur, RuTs, FiFa}.
In the most of the discussions below, we will neglect the 
trivial flat part ${\bf R}^{1,6}$

The explicit metric and other NSNS fields before the Kaluza-Klein reduction 
are given as follows 
\ba
\label{metric}
ds^2&=&d\rho ^2+\f{\rho ^2}{(1+\beta^2\rho^2)(1+q^2\rho^2)}d\vp^2+
\f{1+q^2\rho^2}{1+\beta^2\rho^2}(dy+A_{\vp}d\vp)^2, \no 
A_{\vp}&=&\f{q\rho^2}{1+q^2\rho^2},\ \ \ 
B_{\vp y}\equiv B_{\vp}=-\f{\beta\rho^2}{1+\beta^2\rho^2},\ \ \ 
e^{2(\phi-\phi_0)}
=\f{1}{1+\beta^2\rho^2},  \label{KK}
\ea
where $q, \beta$ are the magnetic parameters which are proportional to the 
strength of two gauge fields and $\phi_0$ is the constant value of the 
dilaton $\phi$ at $\rho =0$. 

It would be useful to note that if $\beta=0$, we get a locally flat
metric 
\ba
\label{metric2}
ds^2=d\rho^2+dy^2+\rho^2(d\vp+qdy)^2.
\ea
This background is globally non-trivial because the angle $\vp$ is 
compactified such that its period is $2\pi$. For example, its geodesic lines
$\vp+qy=$const. are spiral and do not return to the same point  for 
irrational $qR$ if one goes around the circle ${\bf S}^1$.
As we will see later, this geometry rules the D-brane spectrum.

\subsection{Sigma Model Description and Its Free Field Representation}
\hspace{5mm}
At first sight the above background (\ref{KK}) for general $q,\beta$ does 
not seem to be tractable in the description of the 
two dimensional sigma model. 
However, with appropriate T-duality transformations which we will review 
in appendix \ref{Ap:KK}
 (see also \cite{GPR}) one can solve this sigma model in terms of
free fields \cite{BM,SM}\footnote{In our previous paper \cite{TaUe} 
we review the 
path-integral analysis of the Green-Schwarz formalism in the Melvin 
background, while in this paper we use the operator quantization method of 
NS-R formalism in order to construct 
the boundary states.}. In this paper we define the coordinate of 
world-sheet as
$z=\sigma_1+i\sigma_2$ and the derivatives as
$\de=\f{1}{2}(\de_{1}-i\de_{2}), \db=\f{1}{2}(\de_{1}+i\de_{2})$.

The sigma model for the background (\ref{KK}) is given by 
(we show only the bosonic part)
\ba
S&=&\f{1}{\pi\al}\int d^2\sigma\Biggl[\db\rho\de\rho
+\f{(1+q^2\rho^2)}{(1+\beta^2\rho^2)}(\db Y+\f{q\rho^2}{1+q^2\rho^2}
\db\varphi)(\de Y+\f{q\rho^2}{1+q^2\rho^2}\de\vp)\no
&&+\f{\rho^2}{(1+\beta^2
\rho^2)(1+q^2\rho^2)}\db{\varphi}\de{\varphi}
-\f{\beta\rho^2}{1+\beta^2\rho^2}(\db Y\de \varphi-\de Y\db \varphi)
\Biggr]\no
&=&\f{1}{\pi\al}\int d^2\sigma\Biggl[\db\rho\de\rho+\db Y\de Y+
\f{\rho^2}{1+\beta^2\rho^2}(\db \varphi+(q-\beta)\db Y)
(\de \varphi+(q+\beta)\de Y)\Biggr], \no\label{SG1}
\ea
where we have omitted the term of the dilaton coupling for simplicity.
We have also abbreviated the fermion terms since it is easily obtained 
if we use the superfield. 
They are easily incorporated if we use the ${\ca{N}}=1$ world-sheet superfield 
formulation\footnote{To do 
this one has only to replace the derivatives $\de,\ \db$ with $D_{\theta}
=\de_{\theta}+\theta\de,\ 
D_{\bar{\theta}}=\de_{\bar{\theta}}+\bar{\theta}\db$ and a bosonic field 
$X$ with 
${\mb{X'}}(z,\bar{z})
=X(z,\bar{z})+i\theta\psi_{L}(z)+i\bar{\theta}\psi_{R}(\bar{z})+\ddd$.}.

First let us perform the 
T-duality which transforms the field $\vp$ into the new
one $\tvp$ (for more explanations see the appendix \ref{Ap:KK}). 
The result is given by
\ba
S&=&\f{1}{\pi\al}\int d^2\sigma\Bigl[\db\rho\de\rho
+(\db Y+\beta\db \tvp)(\de Y+\beta\de \tvp)
+q(\db \tvp\de Y-\db Y\de\tvp)+\f{1}{\rho^2}\db\tvp\de\tvp\Bigr].\no
\label{SG2}
\ea
After we define the field $Y'$ by
\ba
\label{newval1}
Y'=Y+\beta\tvp,
\ea
we can again take the T-duality along $\tvp$ into $\vp '$. Then we obtain
\ba
S&=&\f{1}{\pi\al}\int d^2\sigma\Bigl[\db\rho\de\rho+\db Y'\de Y'
+\rho^2(\db\vp '+q\db Y')(\de \vp '+q\de Y')\Bigr]. \label{SG3}
\ea
{}From this expression it is easy to see that one can describe the sigma 
model by free fields $X'$ and $\bar{X}'$ which are defined by
\ba
\label{newval2}
X'=\rho e^{i\vp''},~\bar{X}'=\rho e^{-i\vp''},
\ea
where
\ba
\vp ''=\vp '+qY'.
\ea

Here we will examine the relation between the free fields $X',\bar{X}'$ and 
the fields $X=\rho e^{i\vp},\bar{X}=\rho e^{-i\vp}$ which represent the 
original plane ${\bf R^2}\in \mbox{M}_3$ in (\ref{KK}). 
Applying the relation (\ref{rule}) to the 
above two different T-duality transformations\footnote{The same result can be 
obtained by performing the T-duality about
$Y$ once if we regard $\check{\vp}\equiv\vp+qY$ and $Y$ as fundamental fields
in the sigma model. In this subsection we have used the T-duality about $\vp$
for the convenience of the explanation.}, we can obtain
\ba
\de\vp=\de\vp ''-q\de Y-\beta \de Y',\ \ \ \db\vp=\db\vp ''
-q\db Y+\beta \db Y'. \label{p'}
\ea
This shows that the field $\vp$ is rewritten as
\ba
\vp(z,\bar{z})=\vp ''(z,\bar{z})
-qY(z,\bar{z})+\beta\left(Y'_{R}(\bar{z})-Y'_{L}(z)\right),
\ea
where $Y'_{L}(z)~(Y'_{R}(\bar{z}))$ is the left(right)-moving part of $Y'$.
Therefore from this the relation between $X',\bar{X}'$ and 
$X,\bar{X}$ is represented as
\ba
\label{xx}
X(z,\bar{z})=e^{-iqY(z,\bar{z})+i\beta Y'_R(\bar{z})-i\beta Y'_L(z)}
X'(z,\bar{z}).
\ea
It is easy to generalize the above results into those in 
the supersymmetric case since
we can use ${\ca{N}}=1$ world-sheet superfield formulation.
 Then the above equation (\ref{xx}) does hold as a superfield and we
 can also define the free fields ($\psi'_{L,R},\bar{\psi}'_{L,R}$) as the 
partners of $(X',\bar{X}')$.

Since we have the free field representation  
$(Y',\eta'_{L,R})$, $(X',\psi_{L,R})$ and $(\bar{X}',\bar{\psi}_{L,R})$, 
the quantization of the Melvin background can be performed. 
Before that, we have to examine the boundary condition of the field
$Y'$, which is a little subtle analysis in this section. From the
relations (\ref{srule}) one obtains
\ba
\label{tvpj}
&&\de \tvp =-\rho^2\de \vp ''+i\psi'_{L}\bar{\psi}'_{L} \equiv 
i\al j_L,
\no 
&&\db \tvp =\rho^2\db \vp ''-i\psi'_{R}\bar{\psi}'_{R} \equiv - 
i\al j_R ,
\ea
and the conservation law of the above current $j_{L,R}$ follows directly. 
Notice also the useful relation 
\ba
\rho^2\de\vp''=\f{1}{2i}(\bar{X}'\de
X'-X'\de\bar{X}'). \label{r1}
\ea
Then we can define the angular momentum operators 
$\hat{J}_{L},\hat{J}_{R}$ in $(X',\bar{X}')\in{\bf R^2}$ directions as 
follows\footnote{Note the operator product expansions (OPE) 
$j_{L}(z)\de X'(w)\sim \f{1}{2(z-w)}\de X'(w)+\f{1}{2(z-w)^2}X'(w)$, 
$j_{L}(z)\psi'_{L}(w)\sim \f{1}{z-w}\psi'_{L}(w)$, 
$j_{L}(z)\bar{\psi}'_{L}(w)\sim -\f{1}{z-w}\bar{\psi}'_{L}(w)$
and similar results for the right-moving sector. 
Here we have used the relation (\ref{tvpj}), (\ref{r1}) 
and OPE for free fields normalized such that
$X'(z)\bar{X}'(w)\sim -\al\ln (z-w)$ and $\psi'_{L}(z)
\bar{\psi}'_{L}(w)\sim \f{\al}{z-w}$.
Thus we can find that the operators $\psi'_{L,R}$ and 
$\de X',\db X'$ have
charges $\hat{J}_{L,R}=1$ and on the other hand $\bar{\psi}'_{L,R}$ and 
$\de \bar{X}',\db \bar{X}'$ have charges $\hat{J}_{L,R}=-1$.}
\ba
\hat{J}_L=\f{1}{2\pi i}\oint dz\ j_L(z),\ \ 
\hat{J}_R=-\f{1}{2\pi i}\oint d\bar{z}\ j_R(\bar{z}). \label{J}
\ea
Then we can see from (\ref{tvpj}) and (\ref{J}) 
how the boundary condition of $\tvp$ should be twisted 
\ba
\label{tvpperiod}
\tvp(\tau,\sigma+2\pi)=\tvp(\tau,\sigma)-2\pi\al \hat{J},
\ea
where we have defined the total angular momentum operator as $\hat{J}
=\hat{J}_{R}+\hat{J}_{L}$ and 
the new world-sheet coordinates $\tau,\sigma$ as $z=\exp{(\tau+i\sigma)}$.
Moreover notice that the original coordinate $Y$ satisfies
\ba
Y(\tau,\sigma+2\pi)=Y(\tau,\sigma)+2\pi Rw. \label{yw}
\ea
After all from (\ref{newval1}), (\ref{tvpperiod}) and (\ref{yw}) the 
periodicity of the field $Y'$ is given by
\ba
Y'(\tau,\sigma+2\pi)=Y'(\tau,\sigma)+2\pi Rw-2\pi\al\beta \hat{J}. \label{y'w}
\ea

On the other hand, the canonical momentum of $Y$ is 
\ba
P_{Y}&=&\f{1}{2\pi\al}\int d\sigma(q\db\tvp-q\de\tvp+\de Y'+\db Y') \no
&=&q\hat{J}+\f{1}{2}(P'_L+P'_R), \label{py}
\ea
where the first line is obtained from (\ref{SG2}) and (\ref{newval1}).
Therefore from the quantization of $P_Y$ as $P_Y=\frac{n}{R}~(n\in{\bf Z})$ 
the quantized zero modes of $Y'$ are obtained as follows
\ba
\label{zero}
P'_L+P'_R=2(\f{n}{R}-q\hat{J}),\ \ 
P'_L-P'_R=2(\f{Rw}{\al}-\beta\hat{J}).
\ea

Next we turn to the quantization of the free fields $X',\bar{X}'$ and 
$\psi'_{L,R},\bar{\psi}'_{L,R}$. They obey the following 
twisted boundary conditions which can be obtained from (\ref{xx}), (\ref{yw}) 
and (\ref{zero}),
\ba
\label{bc}
X'(\tau,\sigma+2\pi)&=&e^{2\pi i \gamma}X'(\tau,\sigma),\no 
\psi'_{L}(\tau,\sigma+2\pi)&=&
e^{2\pi i \gamma}\psi_{L}'(\tau,\sigma),\ \ \ 
\psi'_{R}(\tau,\sigma+2\pi)=e^{2\pi i\gamma}\psi_{R}'(\tau,\sigma),\\
\label{phase}\no
\mbox{where}~&&~\gamma\equiv qRw+\beta\al(\f{n}{R}-q\hat{J}).
\ea
Note that there are no zero-modes for $X',\bar{X}'$ if $\gamma$
is not an integer. This fact is crucial when we will consider D-branes
in this model later. 
 
The above boundary conditions are similar to those in orbifold 
theories and therefore it is straightforward to perform the mode expansion
and its canonical quantization. 
We summarize these results in the appendix \ref{Ap:Mo}.

\subsection{Mass Spectrum}
\hspace{5mm}
We have explained that the sigma model of the 
Melvin background can be solved in
terms of free fields $(X',\bar{X}',Y')$. Thus it is 
straightforward to compute the mass spectrum of this model in 
the NS-R formalism, which can be obtained from $L_0,\tilde{L}_0$ in 
(\ref{eq6}). If we represent it by using $\hat{N}_{L,R}$ in 
(\ref{NLR}) and $\hat{J}_{L,R}$ in (\ref{JLR}), 
the result \cite{SM, SM2, TaUe} is given by  
\ba
\label{closedmass}
\f{\al M^2}{2}&=&\f{\al}{2R^2}(n-qR\hat{J})^2
+\f{R^2}{2\al}(w-\f{\al}{R}\beta
\hat{J})^2+\hat{N}_R+\hat{N}_L-\hat{\gamma}(\hat{J}_{R}-\hat{J}_{L}),\\
\no
\label{MS1}
&&~~~~~~~~~~~~~~~~~\mbox{where}~~~\hat{\gamma}\equiv\gamma-[\gamma],
\ea
with the level matching constraint
\ba
\label{level}
\hat{N}_R-\hat{N}_L-nw+[\gamma]\hat{J}=0,
\ea 
where $[\gamma]$ denotes the integer part of $\gamma$. 
Moreover the GSO-projection for type II theory restricts the above spectrum, 
which causes a little subtlety \cite{SM, SM2, TaUe}. 
For $2n\leq\gamma<2n+1~(n\in{\bf Z})$ it is the standard type II 
GSO-projection and the allowed spectra are those which give $\hat{N}_{L,R}$ 
the integer values for NSNS-sector. However, for 
$2n+1\leq\gamma<2n+2$ it is the {\sl reversed} one, 
where $\hat{N}_{L,R}$ 
takes half-integer values for NSNS-sector. This fact can be seen 
from the one-loop partition
function $Z(R,q,\beta)$ \cite{SM, SM2, TaUe} of 
the Melvin background by comparing the NS-R formalism with 
the Green-Schwarz formalism, where GSO-projection is not needed. 
In fact the spectrum (\ref{closedmass}) and (\ref{level}) are the same as 
those obtained in the Green-Schwarz formalism \cite{SM, SM2, TaUe}.

Next let us see some interesting symmetries of the partition function 
$Z(R,q,\beta)$. 
{}From the mass spectrum it is easy to show the T-duality relation (see
also appendix A)
\ba
Z(R,q,\beta)=Z(\f{\al}{R},\beta,q).
\ea 
Note that the interchange of $q$ and $\beta$ 
corresponds to that of metric $G_{\vp y}$ and $B$-field $B_{\vp y}$, 
which is the essential part of T-duality
transformation. Furthermore one can see the periodicity of $q$ and $\beta$ 
\ba
Z(R,q,\beta)=Z\left(R,q+\f{2n_1}{R},\beta+\f{2n_2 R}{\al}\right)
~~~~(n_1,n_2\in{\bf Z}).\label{PR}
\ea

\subsection{Supersymmetry Breaking and Relation to Type 0 Theory}
\hspace{5mm}
This Melvin model does not preserve supersymmetry except the case
$(qR,\al\beta/R)\in (2{\bf Z},2{\bf Z})$, which is equivalent to 
the ordinary type II theory in the flat space by the periodicity (\ref{PR}). 
This fact can be seen from the non-vanishing of the partition function 
$Z(R,q,\beta)$. This can also be understood at the level of supergravity. 
Let us assume\footnote{Here we restrict the range of $\beta$
to $0\leq\frac{\beta\al}{R}<2$ because of its periodicity (\ref{PR}).}
$\beta=0$ for simplicity, then all of spin $1/2$ 
fermions which go around the circle ${\bf S}^1$ 
receive the phase 
factor \footnote{This can be also seen if we construct 
the spin field from $\psi'_{L,R}$ with the boundary condition
(\ref{bc}).}$e^{\pm i\pi qR}$. 
Thus only for $qR\in 2{\bf Z}$ the Killing spinors do exist and 
the supersymmetry is preserved. This shows the string theoretic realization 
of Scherk-Schwarz compactification \cite{Ro,AW,KR}. In this kind of
(spacetime) supersymmetry breaking the local supersymmetry is preserved while
 the global supersymmetry is broken \cite{Ro}. 
This fact can be found in the open string spectrum
 on the 
D-brane as we will see in the section 3. For $\beta=0$ and $q\neq 0$ a
D-brane which follows the 
Dirichlet boundary condition for ${\bf S}^1$ has the Bose-Fermi degeneracy
while a D-brane which follows the Neumann boundary condition
does not show the degeneracy.

Thus this model does not preserve supersymmetry in general
and it may be unstable. Indeed
it has tachyons if 
neither $qR$ nor $\al\beta/R$ is an integer \cite{SM}.
In particular, for $(qR,\al\beta/R)
\in (2{\bf Z}+1,2{\bf Z})$, 
the model is identical to type IIA(B) theory twisted by
$(-1)^{F_{S}}\cdot\sigma_{1/2}$ with radius $2R$ as shown in \cite{CG}, 
which is also equivalent to type 0B(A) theory twisted by 
$(-1)^{F_{R}}\cdot\sigma_{1/2}$ with radius $\frac{\al}{R}$ \cite{BG}.
Here the operators $F_{S},F_{R}$ and $\sigma_{1/2}$ represent the
spacetime fermion number, 
the world-sheet right-moving fermion number and the half-shift 
operator on the circle ${\bf S^1}$. 
If we further take the small radius limit $R\to 0$, we obtain
ten-dimensional type 0B(A) string theory \cite{BG} after we perform the 
T-duality. On the other hand, 
if we take the limit $R\to \infty$ with $\beta=0$, 
then the theory is identical
to the ordinal ten dimensional type IIA(B) string theory. 

This equivalence can be generalized to the Melvin background with the 
fractional values of the magnetic parameters \cite{TaUe}.
For the specific fractional 
values\footnote{Here the integers $k$ and $N$ are assumed to be coprime.}
 $qR=\f{k}{N},\ \ \beta=0$ (or $\f{\beta\al}{R}=\f{k}{N},\ q=0$) the 
Melvin model is equivalent to ${\bf Z}_N$ freely acting orbifolds. 
Furthermore if we take the limit $R\to 0$ (or $R\to\infty$), 
those are reduced to
the abelian orbifolds ${\bf C}/{\bf Z}_N$ with a fixed point in type II 
theory (for even $k$) and in type 0 theory (for odd $k$). 
We summarize the result in Table.\ref{orb}.\\
\begin{table}[htbp]
	\begin {tabular}{|c||l||l|}
\hline
  $k$ & type II orbifold (radius) & T-dualized orbifold (radius) 
  \\
\hline
  even & IIA(B)/$\sigma_{1/N}\cdot g\ \ \ \ (NR=\f{N\al}{\ti{R}})$ &
 IIB(A)/$\ti{\sigma}_{1/N}\cdot g\ \ \ \ (\f{\al}{NR}=\f{\ti{R}}{N})$
 \\
  odd  &  IIA(B)/$\sigma_{1/2N}\cdot g \ \ \ (2NR=\f{2N\al}{\ti{R}})$ &
   0B(A)/$\{(-1)^{F_R}\cdot\sigma_{1/2},\ \ti{\sigma}_{1/N}\cdot g\}\ \ \ \ 
  (\f{\al}{NR}=\f{\ti{R}}{N})$ \\
\hline
\end{tabular}
	\caption{The equivalence of the IIA(B) Melvin background for 
	$qR=\f{k}{N},\beta=0$ with freely acting
	orbifolds. The operators $g=\exp{(2\pi i\frac{k}{N}\hat{J})}$, 
        $\sigma_{\frac{1}{N}}$ and $\tilde{\sigma}_{\frac{1}{N}}$ 
        represent the projection operator, $\frac{1}{N}$-shift
	operator in the circle and its dual operator, respectively.}
	\label{orb}
\end{table}

\section{D-branes in Melvin Background}
\setcounter{equation}{0}
\hspace{5mm}
In this section we consider D-branes in Melvin background (\ref{KK}).
In the previous section we have seen that the 
nonlinear sigma model in the Melvin 
background can be exactly solved. 
By applying the T-duality 
in the curved space and by redefining 
appropriate target space variables, the 
nonlinear sigma model can be rewritten 
by the free fields with the nontrivial 
boundary conditions, and we can quantize 
this action in the same way as that in 
the flat space. 

With regard to open strings the quantization process can be also performed 
by the usual method. By changing the boundary conditions of open 
strings we can obtain the various 
D-branes\footnote{For $\beta=0$ case the D-brane systems we discuss 
below are closely related to the D-branes in toroidal compactification
of freely acting orbifolds \cite{An}.} in the Melvin background, while the 
several constraints which can not be seen in the flat space arise due to the 
nontriviality of the Melvin geometry. In the arguments below we define
the D$p$-brane as the D-brane which has $p+1$ Neumann boundary
conditions and $9-p$ Dirichlet boundary conditions in terms of the free
fields $(X',\bar{X}',Y')$, not the original fields $(X,\bar{X},Y)$. The
interpretation of such D-branes in the original coordinate
$(X,\bar{X},Y)$ of the Melvin
background (\ref{KK}) is somewhat complicated as we will discuss later.

One of the interesting results in this section is 
that the D-brane spectra dramatically change
according to whether the parameters in Melvin background $qR$ and 
$\frac{\beta\al}{R}$ take the rational or irrational value. For example 
if $qR$ or $\frac{\beta\al}{R}$ is rational there exist some 
D-branes which are movable in the $(X, \bar{X})\in{\bf R^2}$ 
plane, while if 
both of these parameters take irrational values all 
D-branes are pinned at the origin in the Melvin geometry. 
The second interesting result is that 
some D-branes wrap the nontrivial cycle in the Melvin geometry. 
For example, some 
D1-branes wrap the geodesic lines $\vp+qy=\mbox{constant}$ 
in the Melvin geometry 
without NSNS $B$-field (\ref{metric2}), 
which is the natural result because such a 
configuration gives the minimal mass of the D-brane.\footnote{In the Melvin 
background with $B$-field the analysis becomes 
a little complicated, and we do 
not verified it in this paper.}

However, the most interesting argument about it is that these D1-branes have  
the `spiral' configuration. These wrap  ${\bf S^1}$ in the $Y$ direction, 
however the location of the end point of this 
D1-brane after wrapping one time on ${\bf S^1}$ is not the same as that of the 
start point. In other words 
its end point rotates from the starting point by some 
angle in the $(X,\bar{X})\in R^2$. After wrapping repeatedly, this 
configuration becomes something like a coil wrapping several times on the 
torus. If the parameter $qR$ of the Melvin background is rational, the end 
point of its D1-brane after wrapping several 
times on $S^1$ connects to the 
starting point, while if the parameter is 
irrational its end point never meets 
the starting point and D1-brane wraps on 
the torus infinite times and becomes 
singular because the effective tension of D1-brane diverges. 

There is the other motivation to consider D-branes 
in the Melvin background. As we 
have seen in the previous section or in \cite{TaUe}, by tuning 
magnetic parameters $qR$ and $\frac{\beta\al}{R}$ appropriately 
the various ten dimensional backgrounds 
which we already know can be realized, 
that is, type II, type 0 in flat space and these abelian orbifolds 
${\bf C}/{\bf Z}_N$. Therefore, if we consider D-branes 
in the Melvin background,  
we may understand how the D-branes in the above various backgrounds are 
connected with each other.  

In orbifold theories there exist two types of D-branes which are called 
the fractional D-brane and the bulk D-brane \cite{DoMo, DiDoGo}. 
What we will find is that the fractional D-branes and bulk D-branes 
correspond to the pinned and movable D-branes which we have mentioned, 
respectively. Moreover these D-branes prove to naturally 
correspond not only to the D-branes in type II 
theory in flat space but also 
the bound state of an electric D-brane and a magnetic 
D-brane in type 0 theory 
\cite{KlTs, BeGa2, KlNeSh, BiCrRo, Ga, IM}\footnote{In type 0 
theory two kinds 
of D-branes do appear as we will review later.}. We can verify this by 
the calculation of the tension and the vacuum amplitude etc. by the method 
of the boundary state formalism. 
 
\subsection{Boundary State in Melvin Background}
\hspace{5mm}
In order to investigate D-branes in general 
conformal field theories, it is 
convenient to use the boundary state formalism. The boundary state 
is one way of the representations of D-branes in the closed string Hilbert 
space, and thus from this we can investigate the 
interactions between two D-branes 
and between a D-brane and closed strings. 

{}From previous sections we have already known that 
the action for the Melvin background 
reduces to that written by free fields. Therefore the 
construction of the boundary state is similar to that in the flat space 
(see \cite{VeLi1, VeLi2} and references there in), or more 
precisely to that in orbifold theories \cite{BiCrRo,DiGo,Ta,BiCrRo2}
even though the D-branes which we will
construct have various new intriguing structures. 
Below we will omit the trivial oscillators for ${R^{1,6}}$ directions 
on the world-sheet. 
One can use either the light-cone or covariant formulation.

For the compactified direction $Y'$, the usual 
Neumann or Dirichlet boundary conditions are allowed:
\ba
\label{eq1}
\mbox{Neumann}:&\left\{\begin{array}{ccc}
\de_{\tau} Y'|_{\tau=0}|B\lb=0~&\leftrightarrow&~ 
(\frac{n}{R}-q\hat{J})|B\lb=0,\ \ 
(\beta_m+\tilde{\beta}_{-m})|B\lb=0, \\
(\eta'_{L}-\epsilon\eta'_{R})|_{\tau=0}|B\lb=0~&\leftrightarrow&~
(\eta_r-i\epsilon\tilde{\eta}_{-r})|B\lb=0,
\end{array}\right.\\
\label{eq2}
\mbox{Dirichlet}:&\left\{\begin{array}{ccc}
\de_{\sigma} Y'|_{\tau=0}|B\lb=0~&\leftrightarrow&~
(\frac{Rw}{\al}-\beta \hat{J})|B\lb=0, \ \
(\beta_m-\tilde{\beta}_{-m})|B\lb=0, \\ 
(\eta'_{L}+\epsilon\eta'_{R})|_{\tau=0}|B\lb=0~&\leftrightarrow&~
(\eta_r+i\epsilon\tilde{\eta}_{-r})|B\lb=0,
\end{array}\right.
\ea
where $|B\lb$ is the boundary state and the 
parameter $\epsilon$ takes the value $\pm 1$ which comes 
from open string boundary conditions for world-sheet fermions. 
Note that the zero mode for $Y'$ is given by (\ref{zero}).
For the directions $X', \bar{X}'$ the allowed boundary conditions are 
\ba
\label{eq3}
\mbox{Neumann-Neumann}:&\left\{\begin{array}{ccc}
\de_{\tau} X'|_{\tau=0}|B\lb=0~&\leftrightarrow&~
(\alpha_{m-\gamma}+\tilde{\alpha}_{-m+\gamma})|B\lb=0,\\
\de_{\tau} \bar{X}'|_{\tau=0}|B\lb=0~&\leftrightarrow&~ 
(\bar{\alpha}_{m+\gamma}+\bar{\tilde{\alpha}}_{-m-\gamma})|B\lb=0,\\
(\psi'_{L}-\epsilon\psi'_{R})|_{\tau=0}|B\lb=0~&\leftrightarrow&~
(\psi_{r-\gamma}-i\epsilon\tilde{\psi}_{-r+\gamma})|B\lb=0,\\
(\bar{\psi}'_{L}-\epsilon\bar{\psi}'_{R})|_{\tau=0}|B\lb=0~
&\leftrightarrow&~(\bar{\psi}_{r+\gamma}
-i\epsilon\bar{\tilde{\psi}}_{-r-\gamma})|B\lb=0,
\end{array}\right.\\
\label{eq4}
\mbox{Dirichlet-Dirichlet}:&\left\{\begin{array}{ccc}
\de_{\sigma} X'|_{\tau=0}|B\lb=0~&\leftrightarrow&~
(\alpha_{m-\gamma}-\tilde{\alpha}_{-m+\gamma})|B\lb=0,\\
\de_{\sigma} \bar{X}'|_{\tau=0}|B\lb=0~&\leftrightarrow&~ 
(\bar{\alpha}_{m+\gamma}-\bar{\tilde{\alpha}}_{-m-\gamma})|B\lb=0,\\
(\psi'_{L}+\epsilon\psi'_{R})|_{\tau=0}|B\lb=0~&\leftrightarrow&~
(\psi_{r-\gamma}+i\epsilon\tilde{\psi}_{-r+\gamma})|B\lb=0,\\
(\bar{\psi}'_{L}+\epsilon\bar{\psi}'_{R})|_{\tau=0}|B\lb=0~
&\leftrightarrow&~(\bar{\psi}_{r+\gamma}
+i\epsilon\bar{\tilde{\psi}}_{-r-\gamma})|B\lb=0,
\end{array}\right.
\ea
where $\gamma$ is given by (\ref{phase})\footnote{At this 
stage we can take the more general 
boundary conditions for complex fermions ($\epsilon$ can take U(1) complex 
values and can be unequal to $\epsilon$ in (\ref{eq1}) and (\ref{eq2})). 
However, if we consider the  
${\cal N}=1$ superconformal invariance 
(\ref{supercon}), such boundary conditions are not allowed.}.
Here we have to note that 
the Neumann-Dirichlet or Dirichlet-Neumann  
boundary conditions can be defined 
only if $\gamma$ takes integer or half-integer values. We will discuss these 
special cases in the last of next subsection. We would also like to stress 
that in the above arguments we have defined the Dirichlet or Neumann 
boundary conditions with respect to the free fields $(X',\bar{X}',Y')$.
These boundary conditions are not always equivalent to 
the Dirichlet or Neumann 
boundary conditions with respect to the original fields $(X,\bar{X},Y)$ in the
Melvin sigma model (\ref{SG1}) as we will see later.

{}From these conditions and from (\ref{eq5.5}) and (\ref{eq6}) we can verify 
that the boundary state defined by (\ref{eq1}) $\sim$ (\ref{eq4}) 
satisfies the ${\cal N}=1$ superconformal invariance 
\ba
\label{supercon}
(L_m-\tilde{L}_{-m})|B\lb=0,\no
(G_r+i\epsilon\tilde{G}_{-r})|B\lb=0.
\ea
Moreover from (\ref{eq3}) and (\ref{eq4}) we can 
verify
\ba
\hat{J}|B\lb=(\hat{J}_L+\hat{J}_R)|B\lb=0,
\ea
where the mode expansions of 
$\hat{J}_L$ and $\hat{J}_R$ are given by (\ref{JLR}). 
This shows that these D-branes (NN or DD boundary condition) 
preserve the rotational symmetry on the plane
${\bf R}^2$ as expected\footnote{More precisely, one should take into
account the bosonic zero-mode contribution to the angular momentum 
$\hat{J}_{0}=i\sqrt{\frac{2}{\al}}
(x_{0}\bar{\alpha}_{0}-\bar{x}_{0}\alpha_{0})$ if $\gamma$ takes an 
integer value. However, we can neglect this 
if the D-brane obeys the Neumann-Neumann (NN) boundary condition for 
$(X',\bar{X}')$ or if the D-brane obeys the Dirichlet-Dirichlet (DD) boundary 
condition located at $\rho=0$.
Even if we consider a D-brane with the DD boundary condition and
move it away from the origin $\rho=0$, we can choose $n$ or $w$
such that $(\f{n}{R}-q\hat{J}_{0})|B\lb=0$ or 
$(\f{Rw}{\al}-\beta\hat{J}_{0})|B\lb=0$, respectively. 
We will return this point in section 3.2.}. 

Now we can write down the boundary state explicitly. 
{}From (\ref{eq1}) $\sim$ 
(\ref{eq4}) its explicit form is almost the same as that in the flat space 
\cite{VeLi1, VeLi2}, except the shift of oscillator indices 
by $\hat{\gamma}$ and the GSO-projection. For example, 
the NSNS sector of the boundary state for a D0-brane at $\rho=0$ is
given by
\ba
\label{boundary}
|B,\gamma,\epsilon\lb_{\ss NSNS}=
\exp\left[\sum_{m=1}^{\infty}m^{-1}\beta_{-m}
\tilde{\beta}_{-m}\right]|n,0\lb
\otimes\exp\left[-i\epsilon\sum_{r=1/2}^{\infty}\eta_{-r}\tilde{\eta}_{-r}
\right]|0\lb\nonumber\\
\otimes\exp\left[\sum_{m=0}^{\infty}(m+\hat{\gamma})^{-1}
\alpha_{-m-\hat{\gamma}}\bar{\tilde{\alpha}}_{-m-\hat{\gamma}}
+\sum_{m=1}^{\infty}(m-\hat{\gamma})^{-1}\bar{\alpha}_{-m+\hat{\gamma}}
\tilde{\alpha}_{-m+\hat{\gamma}}\right]|0\lb_{\hat{\gamma}}\nonumber\\
\otimes\exp\left[-i\epsilon\sum_{r=1/2}^{\infty}\{\psi_{-r-\hat{\gamma}}
\bar{\tilde{\psi}}_{-r-\hat{\gamma}}
+\bar{\psi}_{-r+\hat{\gamma}}\tilde{\psi}_{-r+\hat{\gamma}}\}
\right]|0\lb_{\hat{\gamma}+\frac{1}{2}},
\ea
where $\gamma$ is equal to $\frac{\beta\al n}{R}~(n\in {\bf Z})$ as can be seen
from eq.(\ref{phase}) and eq.(\ref{eq2}), and $\hat{\gamma}$ is defined in 
eq.(\ref{MS1})\footnote{We have assumed the specific range
$0<\hat{\gamma}<1/2$. The extension 
of this expression to the other range of $\hat{\gamma}$ and to RR-sector is 
straightforward.}. 
In the above expression we have abbreviated the trivial part which comes from 
the other directions than $X', \bar{X}'$ and $Y'$. Boundary states for 
D-branes which obey other boundary conditions can be obtained similarly. The
total boundary state is given by $|B\lb=|B\lb_{\ss NSNS}\pm |B\lb_{\ss RR}$.
The plus (minus) sign corresponds to a D-brane (an anti D-brane).

Next we have to consider the closed string GSO-projection. 
This is somewhat nontrivial because as we said in the lines below 
(\ref{level}) the GSO-projection for $2n\leq\gamma<2n+1~(n\in{\bf Z})$ is 
the usual projection for type II theory, 
while for $2n+1\leq\gamma<2n+2~(n\in{\bf Z})$ it 
is the projection {\sl with the additional minus sign}. Thus, the 
GSO-invariant boundary state is represented by 
\ba
\label{eq5}
|B\lb_{\ss NSNS}&=&\frac{1}{2\pi R}\sum_{n=-\infty}^{\infty}f_{\gamma}~
\frac{1+(-1)^{[\gamma]}(-1)^{F_L}}{2}
\frac{1+(-1)^{[\gamma]}(-1)^{F_R}}{2}|B,\gamma,+\lb_{\ss NSNS},\no
&=&\frac{1}{2(2\pi R)}\sum^{\infty}_{n=-\infty}f_{\gamma}
\left[|B,\gamma,+\lb_{\ss NSNS}-(-1)^{[\gamma]}
|B,\gamma,-\lb_{\ss NSNS}\right],\no
|B\lb_{\ss RR}&=&\frac{1}{2\pi R}\sum_{n=-\infty}^{\infty}f_{\gamma}~
\frac{1+(-1)^{[\gamma]}(-1)^{F_L}}{2}
\frac{1\mp(-1)^{[\gamma]}(-1)^{F_R}}{2}|B,\gamma,+\lb_{\ss RR},\no
&=&\frac{1}{2(2\pi R)}
\sum^{\infty}_{n=-\infty}f_{\gamma}\left[|B,\gamma,+\lb_{\ss RR}
+(-1)^{[\gamma]}|B,\gamma,-\lb_{\ss RR}\right],~~\left(\mp=
\left\{\begin{array}{ccc}
-&\mbox{for IIA},\no
+&\mbox{for IIB},
\end{array}\right.\right)\\ 
\ea
where $[\gamma]$ is the Gauss symbol 
which picks up the maximal integer part 
of $\gamma$, and constants $f_{\gamma}$ in (\ref{eq5}) are determined by 
the Cardy's condition \cite{Ca}. This is the consistency condition that
the vacuum amplitude between two D-branes computed in the 
closed string sector
by using the boundary state should be equal to the cylinder amplitude of
open string. 

Therefore we would like to calculate the vacuum amplitude by using
the above boundary state. For a D0-brane we obtain the 
following result\footnote{Here we 
have used the explicit form of $L_0$ and $\tilde{L}_0$ in (\ref{eq6}) 
for the closed string propagator. We also have employed the quasi 
periodicity of theta functions (\ref{period}).}\footnote{Note that we 
have divided $\gamma$ into the integer part and the other part, because if 
$\gamma$ takes an integer value the naive calculation of the vacuum amplitude 
by using (\ref{boundary}) diverges. 
This can be seen in the third line of 
eq.(\ref{vv}) if we notice the relation 
$\theta_1(\nu|\tau)=0~(\gamma\in{\bf Z})$ by using (\ref{period}). This 
divergence is due to the reappearance of the zero modes of $X', \bar{X}'$, 
which can be seen from (\ref{bc}) and (\ref{modeex}). 
Therefore, the expression of the boundary state such as 
(\ref{boundary}) is not correct for integer $\gamma$, and we have to 
redefine its bosonic part for $X', \bar{X}'$.} 
\ba
\label{vv}
A&=&\frac{\al}{8\pi R}V_0\sum_{\gamma\in{\bf Z}}~
|f_{\gamma}|^2~\int^{\infty}_0 
ds~(2\pi\al s)^{-4}\exp\left(-\frac{s\al n^2}{2R^2}\right)
(\eta(\tau))^{-12}\no
&&~~~~\times[(\theta_3(0|\tau))^4
-(-1)^{\gamma}(\theta_4(0|\tau))^4
-(\theta_2(0|\tau))^4]\no
&&+\frac{i\al}{8\pi R}V_0\sum_{\gamma\notin{\bf Z}}~(-1)^{[\gamma]}~
|f_{\gamma}|^2~\int^{\infty}_0 
ds~(2\pi\al s)^{-3}\exp\left(-\frac{s\al n^2}{2R^2}\right)
(\eta(\tau))^{-9}(\theta_1(\nu|\tau))^{-1}\no
&&~~~~\times[(\theta_3(0|\tau))^3(\theta_3(\nu|\tau))
-(\theta_4(0|\tau))^3(\theta_4(\nu|\tau))
-(\theta_2(0|\tau))^3(\theta_2(\nu|\tau))],
\ea
where $\tau=\frac{is}{\pi}, \nu=\frac{i\gamma s}{\pi}$.
The volume factor for the time-direction is denoted by $V_0$. By replacing 
$s$ with $\frac{\pi}{t}$ and using the modular transformations 
for theta functions eq.(\ref{TF}) we can obtain the following result
\ba
\label{vac1}
A&=&\frac{\pi\al}{(8\pi R)(2\pi^2\al)^4}V_{0}
\sum_{\gamma\in {\bf Z}}~|f_{\gamma}|^2~\int^{\infty}_{0}
\frac{dt}{t^2}~\exp\left(-\frac{\pi\al n^2}{2R^2 t}\right)
(\eta(it))^{-12}\no
&&~~~~\times[(\theta_3(0|it))^4
-(-1)^{\gamma}(\theta_2(0|it))^4-(\theta_4(0|it))^4]\no
&&+\frac{\pi\al}{(8\pi R)(2\pi^2\al)^3}V_0
\sum_{\gamma\notin {\bf Z}}~(-1)^{[\gamma]}~|f_{\gamma}|^2~\int^{\infty}_{0}
\frac{dt}{t^2}~\exp\left(-\frac{\pi\al n^2}{2R^2 t}\right)
(\eta(it))^{-9}(\theta_1(\gamma|it))^{-1}\no
&&~~~~\times[(\theta_3(0|it))^3(\theta_3(\gamma|it))
-(\theta_2(0|it))^3(\theta_2(\gamma|it))
-(\theta_4(0|it))^3(\theta_4(\gamma|it))].
\ea  

On the other hand the vacuum amplitude can be obtained also from the open 
string one-loop calculation
\ba
\label{vac2}
Z_O=\int^{\infty}_{0}\frac{dt}{2t}
{\rm Tr}_{\ss NS-R}\left[\frac{1+(-1)^F}{2}q^{H_O}\right] \ \ \ \ 
(q\equiv e^{-2\pi t}),
\ea
where we have defined\footnote{The trace includes the factor two due to
the Chan-Paton factor.} 
${\rm Tr}_{\ss NS-R}={\rm Tr}_{\ss NS}-{\rm Tr}_{\ss R}$
; the operator $H_O$ denotes
the open string Hamiltonian\footnote{We are now in the T-dualized space 
where D0-branes live, thus only the winding mode $w$, not the momentum
mode $n$, appears in this open string Hamiltonian.} 
\ba
\label{eq8}
H_O=\al p^2+\al\left(\frac{Rw}{\al}-\beta\hat{J}\right)^2+\hat{N},
\ea 
where $\hat{N}$ and $\hat{J}$ represent the occupation number operator
including the zero point energy ($-1/2$ for NS-sector and $0$ for
R-sector) and the angular momentum generator in ${\bf R}^2$ plane both
of which are the 
open string analog of (\ref{NLR}) and (\ref{JLR}), respectively.  
However, this open string system has two crucially different 
points from closed string one. The first is that there are no tachyons on 
D-branes as we can see from the above equations, while there are in the 
closed string spectrum (\ref{closedmass}). The second is that the indices 
of modes for $X', \bar{X}', Y'$ and their superpartners take 
{\sl the integer or half-integer} (for NS-sector) values because the boundary 
conditions of open strings obey usual Neumann or Dirichlet conditions, 
not twisted ones (\ref{bc}). In particular the mode expansion of $\hat{J}$ 
by open string NS-modes is
\ba
\hat{J}=\frac{i}{\sqrt{2\al}}(x_0 \bar{\alpha}_0-\bar{x}_0\alpha_0)
+\sum^{\infty}_{n=1}\frac{1}{n}(\alpha_{-n}\bar{\alpha}_{n}
-\bar{\alpha}_{-n}\alpha_{n})+\sum^{\infty}_{r=1/2}
(\psi_{-r}\bar{\psi}_{r}-\bar{\psi}_{-r}\psi_{r}),
\ea
where $x_0$ and $\bar{x}_0$ are the zero modes for $X'$ and $\bar{X}'$, 
respectively\footnote{Especially for D0-branes which we are considering here 
the orbital angular momentum part 
$\hat{J}_0=\frac{i}{\sqrt{2\al}}(x_0 \bar{\alpha}_0-\bar{x}_0\alpha_0)$ 
should be neglected due to the absence of the open string zero mode.}.
For R-sector the indices of $\psi$ and $\bar{\psi}$ run integer values, 
and its zero mode contribution $\frac{1}{2}[\psi_0, \bar{\psi}_0]$
should be added. 
{}Therefore we can see that the eigenvalues of $\hat{J}$ take integer 
values for NS-sector and half-integer values for R-sector being 
consistent with the spin-statistics relation. 

Now let us apply the Cardy's condition \cite{Ca} (or open-closed duality). 
By using the Poisson resummation formula 
the open string amplitude (\ref{vac2}) becomes
\ba
\label{vac2'}
A&=&\frac{2\pi\al}{R}
\int^{\infty}_{0}\frac{dt}{2t}(8\pi^2\al t)^{-\frac{1}{2}}\no
&&\times{\rm Tr}_{\ss NS-R}\left[\frac{1+(-1)^F}{2}~q^{\al p^2+\hat{N}}
\sum^{\infty}_{n=-\infty}
\exp\left(-\frac{\pi\al}{2R^2 t}n^2
+2\pi i\frac{\beta\al}{R}n\hat{J}\right)\right].
\ea
By requiring the equality between (\ref{vac1}) and (\ref{vac2'}) we
obtain
\ba
\label{D0tension} 
f_{\gamma}=\left\{\begin{array}{c}
\frac{1}{2}T_0~~~(\gamma \in {\bf Z}),\\
\frac{1}{\sqrt{2}}(\frac{|\sin{\pi\gamma}|}{2\pi^2\al})^{\frac{1}{2}}T_0~~~
(\gamma \notin {\bf Z}),
\end{array}\right.
\ea
where we have defined $T_p=\sqrt{\pi}(2\pi\sqrt{\al})^{3-p}$ and 
$T_p/\kappa$ ($\kappa$ is the gravitational coupling constant) 
is equal to the tension of an ordinary type II Dp-brane 
in flat space. For more general Dp-branes the computations can be performed 
in the same way. Its open string Hamiltonian is the
same as eq.(\ref{eq8}) except the reappearance of the zero modes 
\footnote{The nontrivial relation is the trace formula which comes from the 
open string
zero modes of $X'$ and $\bar{X}'$. This is used for D-branes which obey the 
Neumann-Neumann boundary condition for $X', \bar{X}'$ directions: 
\ba
{\rm Tr}~\exp\left[-2\pi\alpha^{\prime}t p^2+2\pi i\gamma \hat{J}_0
\right]
=\left\{\begin{array}{cc}
(2\sin{\pi\gamma})^{-2}~&~(\gamma\notin {\bf Z}),\no
V_2~(8\pi^2\alpha^{\prime}t)^{-1}~&~(\gamma\in {\bf Z}),
\end{array}\right.
\ea
where
$\hat{J}_0=\frac{i}{\sqrt{2\al}}(x_0 \bar{\alpha}_0-\bar{x}_0\alpha_0)$ is
the orbital angular momentum, and 
$V_2$ is the volume factor for $(X',\bar{X}')\in{\bf R^2}$ plane. 
Such a trace is
familiar in orbifold theories (see for example \cite{Holtak}).}.
The value of $f_{\gamma}$ is given by
\ba
\label{fracten}
f_{\gamma}=\left\{\begin{array}{c}
\frac{1}{2}T_p~~~(\gamma \in {\bf Z}),\\
\frac{1}{\sqrt{2}}(\frac{|\sin{\pi\gamma}|}{2\pi^2\al})^{\mp\frac{1}{2}}
T_p~~~
(\gamma \notin {\bf Z}).
\end{array}\right.
\ea
The sign factors $\mp$ take $-$ for Dp-branes with the Neumann-Neumann 
boundary condition for $X, \bar{X}'$ directions, $+$ for Dp-branes with 
the Dirichlet-Dirichlet boundary condition for those directions. 

The above results for $\gamma\in{\bf Z}$ 
show that a Dp-brane in the Melvin background has 
the same tension as that in the flat space. We would also like to
note that no open string tachyonic modes appear on these Dp-branes.

\subsection{Structure of D-branes for the Rational Parameters}
\hspace{5mm}
As we have said in section two, the nature of the Melvin 
background depends sensitively on whether the (dimensionless) magnetic 
parameters $qR$ and $\f{\beta\al}{R}$ take the rational or irrational values. 
Especially in the former case with $qR=\frac{k}{N}$, $\beta=0$ 
(or $\frac{\beta\al}{R}=\frac{k}{N}$, $q=0$) 
this background is equivalent to 
the freely acting orbifold\footnote{The discussion includes
the $N=1$ case which is equivalent to the type 0 theory
with ${\bf Z}_2$ twist $(-1)^{F_{R}}\cdot \sigma$ \cite{BG}.
For earlier discussions on the D-branes in this model
see \cite{IM,Ko}.}, and under the limit $R\rightarrow 0$ 
($R\rightarrow\infty$) it is reduced to the abelian orbifold 
${\bf C}/{\bf Z}_N$ in type II (for even k) or in type 0 (for odd k) 
\cite{TaUe}. 

In this section we consider D-branes in the Melvin background with the 
rational parameter. As we will see below even for the finite radius 
a single D-brane and $N$ different kinds of D-branes have similar 
properties to a fractional D-brane and a bulk D-brane in orbifold
theories, respectively\footnote{Note that our open string results hold
if either $qR$ or 
$\frac{\beta\al}{R}$ is rational and another is arbitrary. In general the 
closed string theory with these values can not be identified with the freely 
acting orbifold\cite{TaUe}.}. Therefore, from now on, we will often call 
these two types of D-branes in the Melvin background the fractional
D-brane and the bulk D-brane.\\

{\bf Fractional D0-branes}\\

Let us first discuss a D0-brane, which has the Dirichlet boundary
condition in both ${\bf R}^2$ and ${\bf S}^1$ direction. As can be seen
from its boundary state (\ref{boundary}), its behavior depends only on 
$\gamma=\frac{\beta\al}{R}n$. Moreover we can find in (\ref{boundary}) that 
there are no zero-modes in $X',\bar{X}'$ directions unless $\gamma$ takes 
an integer value. We can equally say that this D0-brane can not leave 
from $X'=\bar{X}'=0$ unless $\frac{\beta\al}{R}\in{\bf Z}$, and thus 
the D0-brane is expected to become a 
fractional D0-brane \cite{DoMo,DiDoGo,DiGo} on the orbifolds 
${\bf C}/{\bf Z}_N$ in the limit 
$R\to \infty,~\frac{\beta\al}{R}=\frac{k}{N}$ and $qR\to 0$.
To verify this we begin with the analysis of this orbifold limit.

First we take this limit for the boundary state (\ref{eq5}).
Here we reparametrize the momentum number as 
$n=N\alpha+l~~(\alpha \in {\bf Z},\ l=0,1,\cdots ,N-1)$. In this limit the 
NSNS-sector of the boundary state (\ref{eq5}) becomes
\ba
\!\!\!&&\lim_{R\rightarrow\infty}\frac{1}{2(2\pi R)}\sum_{l=0}^{N-1}
\sum_{\alpha\in{\bf Z}}e^{i\frac{y'}{R}(N\alpha+l)}
f_{\gamma}\left[|B,\frac{kl}{N},+\lb_{\ss NSNS}-(-1)^{k\alpha+[kl/N]}
|B,\frac{kl}{N},-\lb_{\ss NSNS}\right]\no
\no
&=&\left\{\begin{array}{cc}
\displaystyle
\frac{\delta(y')}{2}\sum_{l=0}^{N-1}f'_l
\left[|B,\frac{kl}{N},+\lb_{\ss NSNS}
-(-1)^{[\frac{kl}{N}]}|B,\frac{kl}{N},-\lb_{\ss NSNS}\right]
&~\!\!\!(\mbox{even k}),\\ 
\displaystyle
\!\!\!\frac{\delta(y')}{2}\sum_{l=0}^{N-1}f'_l~ 
|B,\frac{kl}{N},+\lb_{\ss NSNS}
-\!\frac{\delta(y'-\f{\pi R}{N})}{2}\sum_{l=0}^{N-1}(-1)^{[\frac{kl}{N}]}f'_l~ 
|B,\frac{kl}{N},-\lb_{\ss NSNS}
&~\!\!\!(\mbox{odd k}),
\end{array}\right. \label{bso}
\ea
\ba
\label{tension'}
\mbox{where}~~~~~f'_l=\left\{\begin{array}{cc}
\frac{1}{2N}T_0~&~(l=0),\\
\frac{1}{\sqrt{2}N}
(\frac{|\sin{\pi kl/N}|}{2\pi^2\al})^{\frac{1}{2}}T_0~&~
(l\neq 0),
\end{array}\right. \label{bso2}
\ea
and the RR-sector of the boundary state can be obtained in the same way. 
Note that here we extract the $\alpha$ dependent factor from 
the boundary state (\ref{boundary})
which comes only from the zero mode contribution of $Y'$ as
$|n,0\lb=\exp{\left(i\frac{n}{R}y'\right)}|0\lb$. 

For even $k$ this is just the boundary state for a fractional D0-brane in the
type II orbifold ${\bf C}/{\bf Z}_N$
\footnote{Note that in (\ref{bso}) we have not 
used the necessary condition $qR\rightarrow 0$ to realize the orbifolds 
${\bf C}/{\bf Z}_N$ since the boundary state (\ref{bso}) does not 
depend on the value of $q$. This is different from the closed string theory 
which depends on both of the parameters $q, \beta$. We will discuss the 
meaning of this in section 3.4 .}. Indeed we can 
identify the summation over $l$ as the contribution from one untwisted
sector and $N-1$ twisted 
sectors. Moreover the coefficient $f'_l$ in eq.(\ref{tension'}) is $1/N$ 
times of $f_{\gamma}$ in eq.(\ref{D0tension}) and this shows the fractional 
nature of this D-brane explicitly.

\begin{figure} [htb]
\epsfxsize=100mm
\centerline{\epsfbox{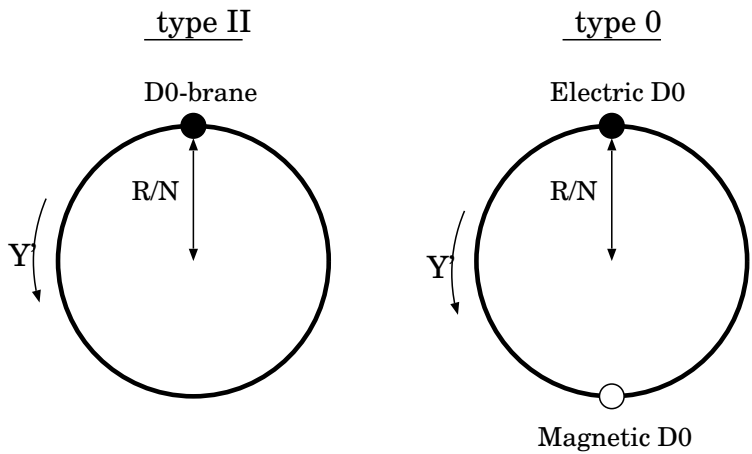}}
\caption{D0-branes}
\label{f1}
\end{figure}

For odd $k$ the boundary state is a little more complicated than for even $k$.
The most important argument is that two kinds of D-branes 
appear each at $y'=0$ (the first term in (\ref{bso})) and at 
$y'=\pi R/N\to \infty$ (the second term in (\ref{bso})). The former 
corresponds to an electric fractional D0-brane and the 
latter a magnetic fractional D0-brane in the orbifold ${\bf C}/{\bf Z}_N$ 
of type 0 theory (see Fig.\ref{f1}). 
The fractional nature of the D-brane can be understood in the same way 
as that in type II theory, and the appearance of two kinds of D-branes is well 
known in type 0 theory \cite{KlTs,BeGa2,BG,IM,Ga,BiCrRo}
\footnote{Of course the two kinds of the corresponding anti D-branes also 
exist even though in this paper we
do not discuss those in the Melvin background.}.  
To explain it let us take the flat type 0 limit $N=1$ (see section 2.3) 
in the boundary state (\ref{bso}). 
Then the untwisted sector ($l=0$) only remains, and if 
we divide it into one with $\delta(y')$ and another with 
$\delta(y'-\pi R)$ those boundary states are given by as follows
\ba
|\mbox{electric}\lb&=&
\frac{T_0}{4}\left( |B,0,+\lb_{\ss NSNS}+|B,0,+\lb_{\ss RR} 
\right),\no 
|\mbox{magnetic}\lb&=&
\frac{T_0}{4}\left( -|B,0,-\lb_{\ss NSNS}+|B,0,-\lb_{\ss RR} 
\right).
\label{0b}
\ea
These are just the boundary states for an electric D0-brane and a magnetic 
D0-brane in the type 0 theory\footnote{Strictly speaking the 
coefficient of the boundary state is a little different. The correct 
coefficient of the boundary state for a type 0 D-brane is $\s{2}$ times as 
large as one in (\ref{0b}). This mismatch comes 
from the fact that the amplitude (\ref{vac2''}) counts the 
open string degrees of freedom 
{\sl on one D-brane}, while in (\ref{bso}) {\sl two D-branes} appear the 
infinite distance away from each other, whose configuration seems singular in 
the uncompactified theory. Thus if we consider {\sl one D-brane} in the 
uncompactified theory (an electric or a magnetic D-brane) we have to multiply 
$\s{2}$ in front of the boundary state. Such a multiplication matches
with the fact that the tension of a type 0 D-brane is $\f{1}{\s{2}}$ times as 
large as that of a type II D-brane \cite{KlTs}.}.
Note that both of the boundary states (\ref{0b}) are invariant under the 
diagonal GSO-projection $(1+(-1)^{F_L+F_R})/2$ which appears in the closed 
string theory of type 0. The appearance of two kinds of D-branes can be 
understood from the fact that the number of massless RR-fields in type 
0A(B) theory is two times as many as those in type IIA(B)
\footnote{Remember that the definition of type 0 theory is the superstring 
theory with the unusual diagonal GSO-projection (see \cite{Po}), and its 
spectrum is given by 
\begin{center}
\begin{tabular}{ccc}
0A~&:&~(NS$+$,NS$+$)~(NS$-$,NS$-$)~(R$+$,R$-$)~(R$-$,R$+$)\no
0B~&:&~(NS$+$,NS$+$)~(NS$-$,NS$-$)~(R$+$,R$+$)~(R$-$,R$-$) 
\end{tabular}
\end{center}
where the sign $\pm$ is denoted by the world-sheet spinor number $(-1)^{F_L}$
and $(-1)^{F_{R}}$.}.

The identification with fractional D-branes can also be shown explicitly by 
examining the vacuum amplitude (\ref{vac2'}). By taking the orbifold limit, 
the summation part of NS-sector in (\ref{vac2'}) becomes
\ba
\label{fracvac1}
&&\lim_{R\rightarrow \infty}\sum_{l=0}^{N-1}\sum_{\alpha\in {\bf Z}}
\exp\left[-\frac{\pi \al}{2R^2 t}(N\alpha+l)^2+2\pi i 
\frac{k}{N}(N\alpha+l)
\hat{J}\right],\no
\!\!&=&\!\!R\int^{\infty}_{-\infty}dp\ 
\exp{\left(-\frac{\pi\al N^2}{2t}p^2\right)}
\sum^{N-1}_{l=0}\exp\left[2\pi i\frac{k}{N}l\hat{J}\right]
\!\!=\!\!
\frac{R}{2\pi\al}(8\pi^2\al t)^{\frac{1}{2}}\frac{\sum^{N-1}_{l=0}~g^l}{N}.
\ea
where $g=\exp[2\pi i \frac{k}{N}\hat{J}]$. Here we have used the fact that 
for NS-sector the eigenvalues of $\hat{J}$ take integers. 
On the other hand for R-sector $\hat{J}$ takes the half-integers, and 
the phase factor $\exp{(2\pi ik\alpha\hat{J})}=(-1)^{k\alpha}$ appears in the 
summation $\sum_{\alpha\in {\bf Z}}$. From this we can see that the amplitude 
in the R-sector becomes zero for odd $k$. Therefore the vacuum amplitude 
(\ref{vac2'}) under the orbifold limit becomes 
\ba
\label{vac2''}
\int^{\infty}_{0}\frac{dt}{2t}
{\rm Tr}_{\ss NS-R}\left[\frac{1+(-1)^F}{2}
\frac{\sum^{N-1}_{l=0}~g^l}{N}~q^{H'_{O}}\right]&~~(\mbox{for even}\ k),\no
\int^{\infty}_{0}\frac{dt}{2t}
{\rm Tr}_{\ss NS}\left[\frac{1+(-1)^F}{2}
\frac{\sum^{N-1}_{l=0}~g^l}{N}~q^{H'_{O}}\right]&~~(\mbox{for odd}\ k),
\ea
where $H'_{O}=\al p^2+\hat{N}$ is the open string Hamiltonian. 
This is just the open string amplitude of a 
fractional D0-brane in type II for even $k$ and that in type 0 for odd 
$k$ (for fractional D-branes in type 0 theory see also \cite{BiCrRo}). 
In particular the absence of R-sector for the type 0 can be understood from 
the fact that there are no fermions on a type 0 D-brane, which can be 
verified by using the boundary state (\ref{0b}). Although the R-sector in type 
0 can appear from the open strings between an electric D-brane and 
a magnetic D-brane, these two D-branes in (\ref{bso}) are infinitely far away 
from each other and its spectrum is neglected.

Let us return to the case of the finite radius $R$. 
{}From the above arguments the D-branes are 
expected to be similar to the fractional D0-branes in the orbifold theories. 
Indeed it is easy to see that the corresponding D0-branes are stuck at the 
point ($X'$=$\bar{X}'$=0) since there are no zero-modes of 
$X',\bar{X}'$ for non-integer $\gamma=\f{\beta\al}{R}n=\f{k}{N}n$ 
(see (\ref{bc}), (\ref{modeex})). 

\begin{figure} [htb]
\epsfxsize=60mm
\centerline{\epsfbox{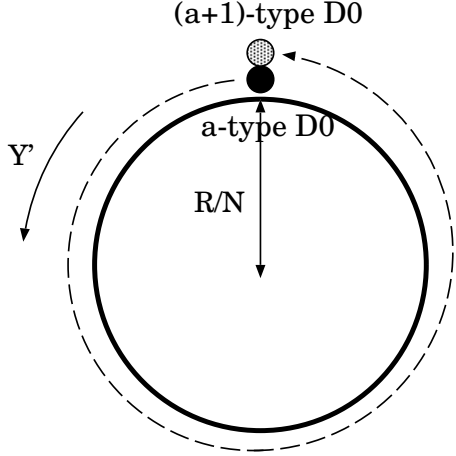}}
\caption{A fractional D0-brane receives the monodromy which changes its 
type.}
\label{f2}
\end{figure}

Up to now we have not considered the $U(1)$ phase in the coefficient 
$f_{\gamma}~(\ref{D0tension})$ of the boundary state (\ref{eq5}). 
We can consider the freedom of the translation of D0-branes in the 
compactified $Y'$ direction
and this effect can be included in the boundary state (\ref{eq5}) by replacing
$f_{\gamma}$ with $f_{\gamma}\exp{(i\frac{n}{R}y'_0)}$ 
($0\leq y'_0<2\pi R$)\footnote{This can be regarded as the 
Wilson line on a D1-brane in the T-dualized picture.}. 
To consider its meaning in the orbifold picture, where it is appropriate to 
regard the radius of ${\bf S^1}$ as $\frac{R}{N}$ 
(see Table.\ref{orb} and eq.(\ref{bso})), we reparametrize $y'_0$ 
as $y'_0=\tilde{y}'_0+\frac{2\pi R}{N}a~
(0\leq\tilde{y}'_0<\frac{2\pi R}{N},~a=0,1,\cdots, N-1)$. Then 
its boundary state is written by
\ba
\label{phase'}
|B_a,\epsilon\lb_{\ss NSNS,RR}
=\sum_{n\in {\bf Z}}e^{i\frac{n}{R}\tilde{y}'_0+2\pi i\frac{a}{N}n}
|B,\gamma,\epsilon\lb_{\ss NSNS,RR}~~~~~~(\epsilon=\pm 1). 
\ea
If we take the orbifold limit $R\rightarrow\infty$, 
the eq.(\ref{phase'}) becomes the eq.(\ref{bso}) except that $\delta{(y')}$ 
and $f'_l$ are replaced by $\delta(y'-\tilde{y}'_0)$ and $f'_l
\exp{(2\pi i\frac{a}{N}l)}$. Then if we remember 
that the fractional D-branes in an orbifold theory are labeled by irreducible 
representations of its discrete group \cite{DoMo,DiDoGo}, we can see that 
these boundary states $(a=0,1\cdots N-1)$ represent the $N$ types of 
fractional D0-branes in the orbifold ${\bf C}/{\bf Z}_{N}$. 
Moreover, we can see that the translation of the D0-brane by 
$\frac{2\pi R}{N}$ in the $Y'$ direction is equivalent to changing the types 
of the fractional D0-brane in the orbifold picture 
since this manipulation is equivalent to $a\rightarrow a+1$ in (\ref{phase'}).
This can be equally said that in the orbifold picture with the radius 
$\frac{R}{N}$ the $N$ types of fractional D0-branes are put at 
the same point $Y'=\tilde{y}'_0$, 
and receive the monodromy to change their types into each other if they
go around ${\bf S^1}$ (see Fig.\ref{f2}).

This observation may be also supported by the mass 
spectrum of open strings between a $a$-type and a $b$-type D0-brane 
(we set $\tilde{y}'_0=0$ for simplicity)
\ba
\label{Mass1}
\al M^2=\f{R^2}{N^2\al}\left(Nw-(k\hat{J}-a+b)\right)^2+\hat{N}.
\ea
For even $k$ the energy due to winding modes cannot vanish unless 
$a-b\in {\bf Z}$ and this shows that there are only $N$ types of D0-branes 
coincident at an arbitrary point in the orbifold picture. 
It would be useful if we could clarify the 
geometrical picture of the D0-brane in terms of the original coordinate
$Y$, not the free field $Y'$, 
of Melvin background (\ref{SG1}). 
This is complicated due to the T-duality transformations in section
2.1. We can speculate that the D0-brane, which is localized in the $Y'$
direction, seems to be extended in the direction $Y$. This may be
regarded as the $\rho=0$ limit of a torus-like D2-brane (see Fig.3) 
which we will discuss later. 

For odd $k$ the energy for NS-sector due to 
winding modes in (\ref{Mass1}) can vanish for the appropriate values of 
$w,~a,~b$ and $\hat{J}$ (for example $w=a=b=\hat{J}=0$), 
while for R-sector it can 
not because $\hat{J}$ takes the half-integer 
values and there remains 
non-zero minimal energy $(\f{R}{2N\al})^2$. 
This shows that a something like a bound system of an electric and a 
magnetic D0-brane exists in the Melvin background, being the finite 
distance $\frac{\pi R}{N}$ away from each other as we have already speculated 
from eq.(\ref{bso})\footnote{Of course these are 
not exactly the same as an electric and a
magnetic D0-brane in type 0 theory, and probably one dissolves into another
because we can not split the boundary state (\ref{eq5}) into each others 
completely except $R\rightarrow\infty$. Moreover this is consistent 
with the fact that there is only one massless RR-field in the Melvin model
 for the finite $R$, which can be seen in (\ref{closedmass}). One can see that 
another RR-field in type 0 theory is massive for the finite $R$, while it 
becomes massless in the uncompactified limit $R\rightarrow\infty$.}. 
These interpretations of D0-branes are shown in Fig.\ref{f1}.

Finally we would like to mention the another orbifold limit $R\to 0$ with 
$qR=\f{k}{N}$ and $\beta=0$. In this case we again obtain the orbifold
${\bf C}/{\bf Z}_N$. However, the boundary state (\ref{boundary}) represents
a bulk D0-brane in this orbifold limit since it depends 
only on $\beta$. Thus one may ask whether there exists a 
D0-brane\footnote{Such a D0-brane, if it exists, should have a 
non-zero winding number $w\neq 0$ and violates the Dirichlet boundary 
condition (\ref{eq2}) for $Y'$. Therefore we must consider the boundary state 
which breaks the $U(1)$ current algebra symmetry.} for 
finite $R$ which is reduced to a fractional D0-brane in the limit $R\to 0$.
Even though this is an intriguing problem, it will be
beyond the scope of the present paper.\\

{\bf Bulk D0-branes}\\

The fractional D0-branes are the most 
fundamental D0-branes in orbifold theories,
and other D0-branes can be constructed by the linear combinations of 
them. In general these D0-branes can not leave from the fixed point
$\rho=0$, while 
if we collect $N$ different types of fractional D0-branes 
they can move as an unit such that $\rho\neq 0$.
The latter can be regarded as another type of 
the D0-brane in the orbifold theories which is called a bulk D0-brane, and 
it is known that the Chan-Paton bundle on this D0-brane obeys the regular 
representation for the discrete group ${\bf Z}_N$ \cite{DoMo}. 

Then it is natural to ask if such a D0-brane 
exists for the finite $R$. 
The answer is yes if the parameter $\frac{\beta\al}{R}$ is 
rational, and the bulk D0-brane in the Melvin background is defined as a 
bound state of $N$ different fractional D-branes whose positions are 
at $N$ different points $Y'=0,\f{2\pi R}{N},\ddd,\f{2\pi R(N-1)}{N}$. 
Its boundary state is given by
\ba
\label{bulk}
|B_{bulk},\epsilon\lb_{\ss NSNS,RR}&=&\sum_{a=0}^{N-1}
|B_a,\epsilon\lb_{\ss NSNS,RR},
\ea
where $|B_a,\epsilon\lb_{\ss NSNS,RR}$ is given by (\ref{phase'}) (here
we set $\tilde{y}'_0=0$). 
The explicit form of the boundary state is obtained in the same way as 
(\ref{bso}) and the result is
\ba
\!\!|B_{bulk},\epsilon\lb_{\ss NSNS,RR}
\!\!=\!\!\frac{T_0/2}{2(2\pi R/N)}\sum_{\alpha\in{\bf Z}}
e^{i\frac{Ny'}{R}\alpha}
\left[|B,0,+\lb_{\ss \!\!NSNS,RR}-\!(-1)^{k\alpha}|B,0,-\lb_{\ss \!\!
NSNS,RR}\right].
\ea
We can see that its boundary state is exactly the same form as 
that for a usual D-brane in type II string (for even $k$) 
or a bound system of an electric and a 
magnetic D-brane away from each other with $\frac{\pi R}{N}$ in
type 0 string (for odd $k$) 
on ${\bf S^1}$ with finite $\frac{R}{N}$ (see Fig.\ref{f1}).  
Note that this boundary state does not have the twisted sectors $l\neq
0$ but picks up only the untwisted sector $l=0$ 
(or $n=N\alpha~(\alpha\in{\bf Z}$)). Thus the bosonic zeromodes $x_0,\bar{x}_0$
indeed exist and the D0-brane can move around such that $\rho\neq 0$.
To complete this argument one should also examine the first condition 
in eq.(\ref{eq2})
carefully since the boundary state with $\rho\neq 0$ has a non-zero
orbital angular momentum $\hat{J}_0$ in the zeromode part. 
Then the condition, which is equivalent to $(w-\f{k}{N}\hat{J}_{0})|B\lb=0$, 
is satisfied 
if $\hat{J}_{0}$ is a multiple of $N$. This requires that a bulk D-brane 
should consist of $N$ fractional D-branes which are located at the $N$
different points
$X'=\rho\ e^{i(\theta+2\pi ka/N)}, \ \ (a=0,1,\ddd, N-1)$ on the plane,
where $\theta$ is an arbitrary constant (see Fig.3). 
In this way we have shown that
a bulk D0-brane exists if $\f{\beta\al}{R}\in {\bf Q}$ for 
any values of parameters\footnote{If the value of $q$ is non-zero, the
vacuum amplitude with non-zero $\rho$ includes an extra term which depends
on $q$. This was recently shown in the case $\beta=0$ ($N=1$) in the 
paper \cite{DuMo}. In the case of rational
$\frac{\beta\al}{R}=\frac{k}{N}$ the open string 
Hamiltonian includes the extra contributions
$\Delta H_{O}=\f{\rho^2}{\pi^2\al}\sin^2(\pi qRw/N+\pi ka/N)$.
This deviation represents the twisted identification due to non-zero $q$
and can be understood from the open string picture.} $q$ and $R$.

Next we consider its vacuum amplitude. 
The result is $N$ times as 
large as (\ref{vac2}), while the second term of the open string 
Hamiltonian (\ref{eq8}) is modified as follows
\ba
\label{bulkvac}
\left(\frac{R(w-k\hat{J})}{N\al}\right)^2\rightarrow\left\{\begin{array}{cc}
(\frac{Rw}{N\al})^2~&~(\mbox{for NS-sector and R-sector with even}~k),\no
\left(\frac{R(w+1/2)}{N\al}\right)^2~&~(\mbox{for R-sector with odd}~k).
\end{array}\right.
\ea
This is because the angular momentum operator $\hat{J}$ takes integer 
(half-integer) values in NS-sector (R-sector) and we can erase the effect of 
$\hat{J}$ by the shift of $w$. 
{}From this Hamiltonian we can see that there are no tachyonic modes 
though in the closed string sector there are in general. 
If we take the orbifold limit 
$R\rightarrow\infty$, the vacuum amplitude is $N$ times as large as 
(\ref{vac2''}) without the orbifold projection 
$\sum^{N-1}_{l=0} g^l/N$, 
which is the same as that in the flat type II theory for even $k$ or type 0
theory for odd $k$. 

\begin{figure} [htb]
\epsfxsize=150mm
\centerline{\epsfbox{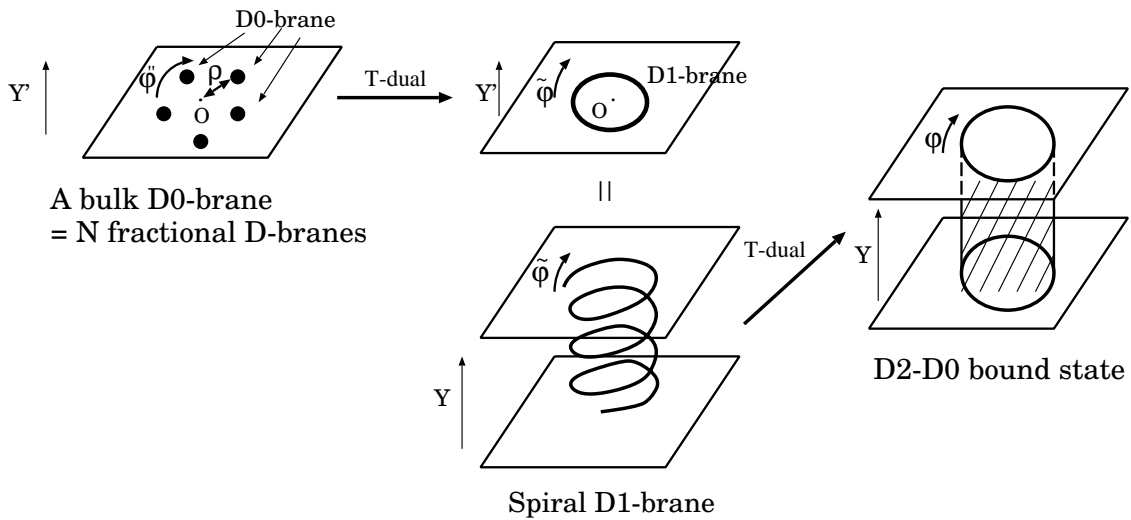}}
\caption{A bulk D0-brane can be regarded as a D2-D0 bound state in the
Melvin background (\ref{KK}).}
\label{f4}
\end{figure}

Finally we would also like to see how this D0-brane looks like
in terms of the original Melvin coordinate $(X,\bar{X},Y)$.
Remember that the coordinates $(X',\bar{X}',Y')$ and $(X,\bar{X},Y)$
are related with each other by performing the T-duality twice as
$\vp''\to \tvp\to \vp$ (see section 2.1). After the first T-duality procedure on $\vp''$,
the D0-brane located at $Y'=\mbox{fixed},\ 
\rho=\mbox{fixed}(\neq 0)$ and $\vp''=\mbox{fixed}$
will be changed into a D1-brane wrapped on the circle $0\le \tvp<2\pi\al$. 
If we see this in terms of $(Y,\tvp)$, the D1-brane can be viewed as 
a spiral D-string wrapped $N$ times on the circle $0\le \tvp<2\pi\al$ and 
$k$ times on the circle $0\le Y<2\pi R$ because of the relation 
$Y=Y'-\beta\tvp$ in (\ref{newval1}) . If we take the second T-duality on
$\tvp$, then we obtain a D2-brane wrapped on the torus 
${\bf T}^2\ni (Y,\vp)$ (see Fig.\ref{f4}). This D2-brane should be regarded
as {\sl a bound state of $k$ D2-branes and $N$ D0-branes} as can be seen from
the winding numbers of the spiral D-string. Note that this is consistent
with the asymptotic value of $B$-field in (\ref{KK}) 
$B_{\vp}=-\frac{1}{\beta}=-\f{N\al}{kR}$ 
if $\beta\rho\gg 1$. {}From this value $B_{\vp}$ 
we can show that if we consider the 
world-volume theory of a D2-brane on a torus ${\bf T}^2$ in the Seiberg-Witten 
limit \cite{SeWi}, we will obtain the theory on a noncommutative torus
with the noncommutativity parameter $\theta=\f{k}{N}$.\\

{\bf D1-branes Wrapped on ${\bf S}^1$}\\

As we have seen, D0-branes in the Melvin background with the rational
parameter are very similar to those in ${\bf Z}_N$ orbifolds. 
On the other hand, a D1-brane wrapped on ${\bf S}^1$ has an interesting
structure which can not be explained clearly 
from the viewpoint of orbifold theories
even though a D1-brane is formally 
transformed into a D0-brane by T-duality. 

Here we consider the Melvin background with the rational parameter 
$qR=\frac{k}{N}$. The boundary state of a D1-brane can be constructed in the 
same way as that of a D0-brane. 
Then a single D1-brane is again pinned at the origin $\rho=0$ (fixed point) in 
(\ref{KK}). However if we consider $N$ D1-branes so that boundary state 
includes only the restricted winding sectors $w\in N{\bf Z}$, this bound 
system (`bulk' D1-brane) can move around in the plane ${\bf R^2}$ in the same 
way as in the previous case of D0-branes.

\begin{figure} [htb]
\epsfxsize=80mm
\centerline{\epsfbox{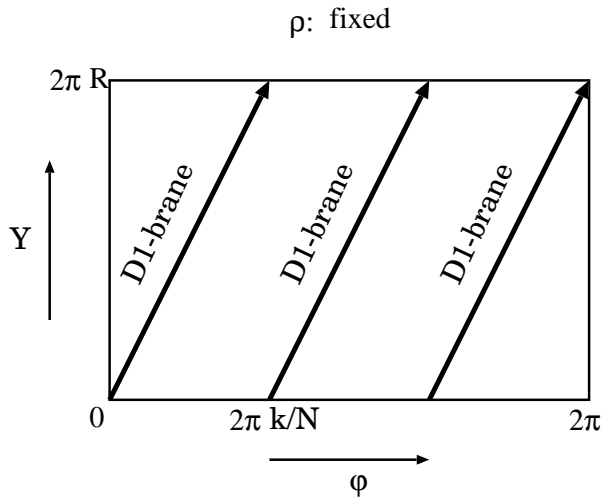}}
\caption{Spiral D1-branes}
\label{f3}
\end{figure}

This behavior can be also explained geometrically as follows. 
Let us set $\beta=0$ for simplicity and assume that a single D1-brane is
placed at $\rho\neq 0$.  
Though this D1-brane obeys the Dirichlet 
boundary condition along $(X',\bar{X}')\in{\bf R}^2$,
in the original coordinate $(X,\bar{X})$ 
it is rotated by the angle $\f{2\pi k}{N}$  
if it goes around ${\bf S}^1$ once, as shown in eq.(\ref{xx}). Thus it
should wind $N$ times around ${\bf S}^1$ in order to move around on the 
${\bf R^2}$ plane (see Fig.\ref{f3}). 
It is also useful to note
that the geodesic lines along ${\bf S}^1$ are given by $\vp+qy=
\mbox{constant}$ from (\ref{metric2}) and agrees with the world-volume of the 
D1-brane. This is a good 
tendency since it will minimize the mass of the D-brane\footnote{For
non-zero value of $\beta$ the world volume of D1-brane seems to be the
same $\vp+qy=\mbox{constant}$, which can be seen by examining the
T-duality procedures as before. This is not exactly 
coincident with the geodesic line $\vp+(q\pm\beta)y=\mbox{constant}$ 
for (\ref{KK}).}.

The analysis of the orbifold limit $R\to 0$ with $qR=\f{k}{N}$ and $\f{\beta
\al}{R}\to 0$ can be performed in the same way as before. 
A single D1-brane (below we reuse the coordinate $(X',\bar{X}',Y')$) 
which is sticked at the fixed point is identical to a fractional D0-brane in 
the limit by T-duality. On the other hand, the D1-brane winding $N$
times around ${\bf S}^1$
corresponds to a bulk D0-brane which is made of $N$ fractional D0-branes in 
the orbifold ${\bf C}/{\bf Z}_N$. 

Let us consider the vacuum amplitude for a `bulk' D1-brane.
The momentum part of the open string Hamiltonian 
(the T-dual picture of (\ref{eq8})) is given by
\ba
\left(\frac{n-k\hat{J}}{NR}\right)^2\rightarrow\left\{\begin{array}{cc}
(\frac{n}{NR})^2~&~(\mbox{for NS-sector and R-sector with even}~k),\no
(\frac{n+1/2}{NR})^2~&~(\mbox{for R-sector with odd}~k).
\end{array}\right.
\ea
Note that for odd $k$ case the obtained D1-brane is really a system which 
consists of an electric `bulk' D1-brane and a magnetic `bulk' D1-brane. 
They wind $N$ times around ${\bf S}^1$ and there is also a ${\bf Z}_2$ 
Wilson line on either D1-brane. 

In this way we have found that a D1-brane wrapped on ${\bf S}^1$
in the Melvin background has the spiral structure (see Fig.\ref{f3}). 
As can be seen from the 
open string Hamiltonian there are again no tachyonic modes even though in the
closed string sector there are tachyons in general.\\

{\bf Other Dp-branes}\\

The analysis of other D-branes can be done in the same way. 
We have only to 
consider the three dimensional Melvin geometry out of the ten dimensional 
spacetime. Then both a D3-brane and a D2-brane which are extended in the 
$X', \bar{X}'$ directions are also allowed (see the Table.1). 
Each of these becomes 
a `fractional'\footnote{Usually in orbifold theories the 
Dp-brane ($p\geq 1$) 
is not called fractional because its tension is not divided by $N$, while in 
this paper `fractional' Dp-branes are ones which have the 
open string vacuum 
amplitude with the orbifold projection in the orbifold limit 
such as (\ref{vac2''}) in order to 
distinguish this from the `bulk' Dp-brane without the orbifold
projection.} D2-brane on the space ${\bf C}/{\bf Z}_N$ in the 
orbifold limit 
(for D3-branes we take a T-duality transformation in the $Y'$ direction). 
This kind of D2-brane has the same tension as an ordinary 
D2-brane in almost the same way as in orbifold theories \cite{Holtak}. 

For the rational case there exists another D-brane
which obeys the Neumann-Dirichlet or Dirichlet-Neumann boundary conditions 
for $X',\bar{X}'$ directions as we mentioned below (\ref{eq4}). 
For bulk D-branes we can define such boundary conditions because on such 
D-branes the parameter $\gamma$ which appears in the boundary states takes 
an integer value\footnote{Remember that bulk D-branes do 
not have twisted sectors.}. Especially note that these D-branes can not be 
defined as bound systems of $N$ kinds of fractional D-branes such as 
(\ref{bulk}). 

Such an example also appears for 
$\frac{\beta\al}{R}=\frac{k}{N}$ (or $qR=\frac{k}{N}$) with even $N$ 
and odd $k$. In this background we can consider a D-brane with the
untwisted sector $\hat{\gamma}=0$ and {\sl one twisted sector 
$\hat{\gamma}=\frac{1}{2}$}, which is just the condition to allow the mixed 
boundary conditions (see below (\ref{eq4}))\footnote{Of course we can
define the D-brane 
with Neumann-Neumann or Dirichlet-Dirichlet boundary conditions 
for $X',\bar{X}'$ directions by the linear combination 
of $\frac{N}{2}$ kinds of fractional D-branes, 
while we can not construct the D-branes with mixed boundary conditions from 
the fractional D-branes.}. 

Here we have obtained the several D-branes in Melvin background, thus we 
summarize these results in the Table \ref{D-brane1}.\\

\begin{table}[htbp]
	\begin {tabular}{|c|c||c|c|c|c||c|}
\hline
  $Y'$ & $X',\bar{X}'$ & Existence & Mobility & Tension & Form 
  & Orbifold Limit \\
\hline
    D & DD & all & Pinned & $T_{0}$ & a D0-brane & fractional D0 \\
    D & DD & $\f{\beta\al}{R}=\f{k}{N}$ & Movable & $NT_{0}$ & 
    $N$ D0-branes & bulk D0 \\
    D & DN & $\f{\beta\al}{R}=\f{k}{N}$ & Movable & $NT_{1}$ & 
    $N$ D1-branes & bulk D1 \\
    D & NN & all & --- & $T_{2}$ & a D2-brane & `fractional' D2 \\
\hline
    N & DD & all & Pinned & $T_{1}$ & a D1-brane & fractional D0 \\
    N & DD & $qR=\f{k}{N}$ & Movable & $NT_{1}$ & $N$ spiral D1-branes
     & bulk D0 \\
    N & DN & $qR=\f{k}{N}$ & Movable & $NT_{2}$ & $N$ spiral D2-branes 
     & bulk D1  \\
    N & NN & all & --- & $T_{3}$ & a D3-brane & `fractional' D2 \\
    
\hline
\end{tabular}\\
	\caption{D-brane spectrum in the Melvin background. Here we
	define the boundary condition of open strings in terms of free
	fields $(X',\bar{X}',Y')$. Note that for 
	odd $k$ we regard a D$p$ in the above as a bound state of 
an electric D$p$ and a magnetic D$p$. }
	\label{D-brane1}
\end{table}


\subsection{Structure of D-branes for the Irrational Parameters}
\hspace{5mm}
As we have already mentioned, 
the D-brane spectrum for the irrational parameters
is remarkably different from the previous case of rational parameters.
A single D-brane can exist only at the origin $\rho=0$ in (\ref{metric}) 
as before. However if one wants to put D-branes at the other points 
$\rho\neq 0$, then one must prepare infinitely many D-branes, which can
be verified by calculating its tension in the same way as before.   
This fact matches the intuitive picture of the irrational case as a
large $N$ limit of the orbifold ${\bf C}/{\bf Z}_N$ \cite{TaUe}. 
For example, let us consider a D1-brane wrapped
on ${\bf S}^1$. In the irrational case the D1-brane along the 
geodesic line cannot return to 
the original point even if it goes around ${\bf S}^1$
any times. Thus the system will be a sort of a `foliation' of the
cylinder $\rho=\mbox{constant}$ and may be like a 2-brane. Even though we 
cannot answer whether such a system
can really exist, we can say that the D-brane spectrum for the 
irrational case is more restricted than that for the rational case.

\subsection{World-Volume Theory}
\hspace{5mm}
In the above we have constructed boundary states of various D-branes and
have seen their geometric interpretations. Here we would like to briefly 
discuss the 
world-volume theory on these D-branes. 

First note that the boundary states
which have the Neumann (or Dirichlet) 
boundary condition in the $Y'$ direction 
depend only on the parameter $qR$ 
(or $\frac{\beta\al}{R}$) and not on another. This fact is the reason why 
a D-brane in the background where {\sl one of the parameters is rational}
is treated as if it was a fractional D-brane, even though another
parameter is irrational and this background is not equivalent to any
freely acting orbifolds.

This fact will also lead
to the intriguing property that the mass spectrum of open string which 
obeys the Dirichlet boundary condition along ${\bf S}^1$
can {\sl show the Bose-Fermi degeneracy}\footnote{Note 
also that this fact may be correct
only up to the one-loop order on the string coupling constant. This is 
because the closed one-loop (open two loop) amplitude between two D-branes 
contains the two point correlation function on the torus, and in this 
amplitude both parameters $qR$ and $\frac{\beta\al}{R}$ appear 
(see \cite{SM,TaUe}). } for $\beta=0$ and $q\neq 0$
even if the {\sl supersymmetry is completely
broken} in the closed string sector. This mysterious phenomenon can be
a relic of the supersymmetry in type II theory, which is spontaneously broken
\cite{Ro} by the non-zero value of $q$. In other words the Bose-Fermi 
degeneracy only on the point-like D-brane along ${\bf S}^1$
implies the remained local supersymmetry, while the degeneracy does not 
occur on the D-brane wrapped on ${\bf S}^1$ because of the lack of global
supersymmetry.

Next let us consider the relation to the quiver gauge theory 
\cite{DoMo}. In particular we take a D0-brane as an example.
Let us remember the open string mass spectrum (\ref{Mass1}).
If we take the orbifold limit $R\to \infty$,
 then it is easy to see that the ${\bf Z}_N$ projection of quiver theory
\ba
k\hat{J}-a+b=Nw\in N{\bf Z}, 
\ea
appears, which restricts the spectrum to 
$g~(=\exp{[2\pi i\frac{k}{N}\hat{J}]})=\exp{[2\pi i\frac{a-b}{N}]}$. 
Thus the world-volume theory on D0-branes in the Melvin background can be 
regarded as a deformation of the quiver theory whose ${\bf Z}_N$
projection is softened. One 
can also see that the opposite limit $R\to 0$ leads to the mass spectrum of
an ordinary D0-brane in type II theory. Thus for a finite radius the 
world-volume theory is regarded as an interpolation between them. In 
particular the massless fields are the same as those of the quiver gauge 
theory of the orbifold ${\bf C/Z}_N$.

\section{D-branes in Supersymmetric Higher Dimensional Models}
\setcounter{equation}{0}
\hspace{5mm}
The closed string backgrounds we have discussed above 
do not preserve any supersymmetry in general.
If we would like to discuss D-brane charges, it will be more desirable
to consider those in supersymmetric backgrounds. Therefore in this section
we investigate D-branes in the higher dimensional 
generalization of the Melvin 
model \cite{TaUe,RuTs}, which includes the supersymmetric background. 
This model has a background of the form
$\mbox{ \bf M}_5\times {\bf R}^{1,4}$, where $\mbox{\bf M}_5$ 
is a ${\bf S}^1\ni Y$
 fibration over ${\bf R}^2\times {\bf R}^2\ni 
 (X^1,\bar{X}^1,X^2,\bar{X}^2)$.
 Correspondingly this model has four magnetic parameters 
 $q_1,q_2,\beta_1$ 
 and $\beta_2$, and the explicit metric of this model is given 
 in the paper \cite{TaUe}. The four of eight (light-cone gauge) 
 spinor fields in the Green-Schwarz formulation do not 
 suffer from the phase factor when $\sigma$ is shifted by $2\pi$
  iff $q_1=q_2,\beta_1=\beta_2$ or $q_1=-q_2,\beta_1=-\beta_2$. 
  Therefore
 we can conclude that in these specific cases 
 half of thirty two supersymmetries will be preserved. Indeed we can also 
show that the partition function does vanish for these 
cases \cite{TaUe}. 
 From the supergravity viewpoint, we can see this as follows. For 
 simplicity, let us set $\beta_1=\beta_2=0$. Then 
 if the spinor fields go around the circle
 ${\bf S}^1$, they obtain the phase 
 $e^{i\pi(\pm q_1\pm q_2)R}$.
 Thus if $q_1=q_2$ or $q_1=-q_2$, there are sixteen Killing spinors. 
 This special
case $q_1=\pm q_2,\beta_1=\beta_2=0$ can be regarded as 
the `$9-11$' flip of the
supersymmetric F5-brane \cite{GS}.
  We would also like to mention that 
 similar arguments can also be generalized into much higher dimensional
 backgrounds $\mbox{\bf M}_7$ (F3-brane), $\mbox{\bf M}_9$ (F1-brane) 
 \cite{TaUe,RuTs}. The results on the behavior of 
 D-branes discussed below
will also be applied to these more generalized cases without any serious 
modifications.

\subsection{Sigma Model Description and Mass Spectrum}
\hspace{5mm}
If we consider the higher dimensional generalizations of (\ref{SG1}) 
its world-sheet action is given as follows :
\ba
S&=&\f{1}{\pi\al}\int d^2\sigma\Bigl[\db\rho\de\rho+\db r\de r+
\rho^2\db\check{\vp}\de\check{\vp}+r^2\db\check{\theta}\de\check{\theta}\no
&&+(1+\beta_1^2\rho^2+\beta_2^2 r^2)^{-1}(\db Y+\beta_1\rho^2\db\check{\vp}
+\beta_2 r^2 \db\check{\theta})(\de Y-\beta_1\rho^2\de\check{\vp}
-\beta_2 r^2 \de\check{\theta})\Bigr],
\ea
where we have defined $\check{\vp}=\vp+q_1Y,\ \check{\theta}=
\theta+q_2Y $.
After T-duality of $Y$ into $\ti{Y'}$\footnote{In this section we take
the different process of T-duality from that in section 2.}, we obtain
\ba
\label{superaction}
S=\f{1}{\pi\al}\int d^2\sigma\left[\db\rho\de\rho+\db r\de r+\db\ti{Y'}
\de\ti{Y'}+\rho^2\db \vp '' \de \vp ''+
r^2\db \theta '' \de\theta '' \right],
\ea
where $\vp ''\equiv\check{\vp}-\beta_1\ti{Y'},\ \ \theta''
\equiv\check{\theta}-\beta_2\ti{Y'}$ are the higher dimensional 
generalization\footnote{This correspondence can be shown if one notes that
the relations (\ref{rule}) are rewritten as 
$-\de\ti{Y'}=\de Y-\beta_1\rho^2\de\vp''-\beta_2 r^2\de\theta''$ and
$\db\ti{Y'}=\db Y+\beta_1\rho^2\db\vp''+\beta_2 r^2\db\theta''$.}
of $\vp''$ which appeared in eq.(\ref{p'}). The T-dual of $\ti{Y'}$ is equivalent
to $Y'$ in eq.(\ref{SG3}). The zero-mode of $Y'$ is quantized
in the same way as (\ref{zero})
\ba
P'_{L}+P'_{R}=2 \left(\f{n}{R}-q_1\hat{J}_1-q_2\hat{J}_2 \right),\ \ \ 
P'_{L}-P'_{R}=2 \left(\f{Rw}{\al}-
\beta_1 \hat{J}_1-\beta_2 \hat{J}_2 \right),
\ea
where the angular momenta 
$\hat{J}_1=\hat{J}_{1L}+\hat{J}_{1R}$ and $\hat{J}_2=\hat{J}_{2L}+
\hat{J}_{2R}$ are defined for each
${\bf R}^2$ as were done in eq.(\ref{J}).

{}From (\ref{superaction}) 
the bosonic fields $X^{1'}\!=\!\rho e^{i\vp''},\ \ X^{2'}\!=\!
r e^{i\theta''}$ and
their superpartners $\psi^{1'}_{L},\psi^{1'}_{R},
\psi^{2'}_{L},\psi^{2'}_{R}$ become free fields and they obey
the following boundary conditions for each $i=1,2$
\ba
X^{i'}(\tau,\sigma+2\pi)&=&e^{2\pi i \gamma_i}X^{i'}(\tau,\sigma),\no
\psi^{i'}_{L}(\tau,\sigma+2\pi)
&=&e^{2\pi i \gamma_i}\psi^{i'}_{L}(\tau,\sigma), \ \ \ \ 
\psi^{i'}_{R}(\tau,\sigma+2\pi)=e^{2\pi i\gamma_i}\psi^{i'}_{R}
(\tau,\sigma),\\
\no
\mbox{where}&&\hat{\gamma_i}\equiv\gamma_i-[\gamma_i],\ \ \ \gamma_i\equiv
q_i Rw+\beta_i\al(\f{n}{R}-q_1 \hat{J}_1 -q_2 \hat{J}_2). \label{gam}
\ea

Then we can obtain the mass spectrum as follows
\ba
\f{\al M^2}{2}&=&\f{\al}{2R^2}(n-q_1 R\hat{J}_1-q_2 R\hat{J}_2)^2
+\f{R^2}{2\al}(w-\f{\al}{R}\beta_1\hat{J}_1-\f{\al}{R}\beta_2\hat{J}_2)^2
\no
&&+\hat{N}_R+\hat{N}_L-
\sum_{i=1}^{2}\hat{\gamma_i}(\hat{J}_{Ri}-\hat{J}_{Li}),
\ea
with the level matching constraint
\ba
\hat{N}_R-\hat{N}_L-nw+\sum_{i=1}^{2}[\gamma_i]\hat{J}_i=0,
\ea 
where $\hat{N}_{L,R}$ is defined in the same way as (\ref{NLR}).
{}From the above expression we can find the T-duality symmetry
$q_i\lr \beta_i,~R\lr \frac{\al}{R}$ if $q_1\beta_2=q_2\beta_1$. Note that the 
supersymmetric model ($q_1=\pm q_2$, $\beta_1=\pm\beta_2$) 
satisfies this condition.

It is also easy to see that if $n_1+n_2,\ n_1-n_2\in 2{\bf Z}$,
we obtain the following periodicity
\ba
Z(R,q_1,q_2,\beta_1,\beta_2)=Z(R,q_1+\f{n_1}{R},q_2+\f{n_2}{R},
\beta_{1},\beta_{2}), \label{TDH}
\ea
and the periodicity for $\beta_i$ can be also obtained by the T-duality.

\subsection{Supersymmetric D-branes}
\hspace{5mm}
We can consider D-branes in the higher dimensional Melvin model in the 
same way as in section three. The quantization of open strings can 
be performed because the action is already rewritten by free fields and we 
can construct the boundary states of D-branes. Its explicit form is
almost the same as in eq.(\ref{boundary}) and eq.(\ref{eq5}). 
By using such boundary states we can obtain the vacuum amplitude, for 
example, for a D0-brane as follows
\\

\begin{eqnarray}
A&=&\frac{\pi\al V_0}{8\pi R}\int^{\infty}_{0}\frac{dt}{t^2}\no
&&\times\Biggl[\Bigl\{
\sum_{\gamma_1\in{\bf Z},\gamma_2\in{\bf Z}}
\frac{|f_{\gamma}|^2}{(2\pi^2\al)^4}(\eta(it))^{-12}
+\sum_{\gamma_1\notin{\bf Z},\gamma_2\in{\bf Z}}
\frac{(-1)^{[\gamma_1]}|f_{\gamma}|^2}{(2\pi^2\al)^3}
(\eta(it))^{-9}(\theta_1(\gamma_1|it))^{-1}\no
&&~~~~~~~~~~~~~~~+\sum_{\gamma_1\in{\bf Z},\gamma_2\notin{\bf Z}}
\frac{(-1)^{[\gamma_2]}|f_{\gamma}|^2}{(2\pi^2\al)^3}(\eta(it))^{-9}(\theta_1(\gamma_2|it))^{-1}\no
&&~~~~~~~~~~~~~~~+\sum_{\gamma_1\notin{\bf Z},\gamma_2\notin{\bf Z}}
\frac{(-1)^{[\gamma_1]+[\gamma_2]}|f_{\gamma}|^2}
{(2\pi^2\al)^2}(\eta(it))^{-6}
(\theta_1(\gamma_1|it))^{-1}(\theta_1(\gamma_2|it))^{-1}\Bigr\}\no
&&~~~~~~~~~~~~~~~~~~~~\times\exp\left(-\frac{\pi\al}{2R^2 t}n^2\right)
(\theta_1(\frac{\gamma_1+\gamma_2}{2}|it))^2
(\theta_1(\frac{\gamma_1-\gamma_2}{2}|it))^2\Biggr],\label{VAH}
\end{eqnarray}
where $\gamma_i=\frac{\beta_i\al}{R}n$, and the summations should be 
performed about $n\in {\bf Z}$ such that
the conditions indicated below the symbol $\sum$ are satisfied. Note also 
that we have transformed the result obtained in the NS-R formulation 
into that in the Green-Schwarz formulation using the formula (\ref{TF2}).

Such vacuum amplitude should be consistent with the Cardy's condition and
this determines the coefficient $f_{\gamma}$ 
which appears as in eq.(\ref{eq5}). The
result for a general Dp-brane is given by
\ba
\label{superfracten}
f_{\gamma}=\left\{\begin{array}{cc}
\frac{1}{2}T_p~&~(\gamma_1\in {\bf Z},\gamma_2\in {\bf Z}),\\
\frac{1}{\sqrt{2}}(\frac{|\sin{\pi\gamma_1}|}{2\pi^2\al})^{\mp\frac{1}{2}}
T_p~&~(\gamma_1\notin {\bf Z},\gamma_2\in {\bf Z}),\\
\frac{1}{\sqrt{2}}(\frac{|\sin{\pi\gamma_2}|}{2\pi^2\al})^{\mp\frac{1}{2}}
T_p~&~(\gamma_1\in {\bf Z},\gamma_2\notin {\bf Z}),\\
(\frac{|\sin{\pi\gamma_1}|}{2\pi^2\al})^{\mp\frac{1}{2}}
(\frac{|\sin{\pi\gamma_2}|}{2\pi^2\al})^{\mp\frac{1}{2}}
T_p~&~(\gamma_1\notin {\bf Z},\gamma_2\notin {\bf Z}),
\end{array}\right.
\ea
where the sign factor $\mp$ in the exponent of $|\sin\pi\gamma_i|~(i=1,2)$ 
takes $-$ for D-branes with Neumann-Neumann boundary condition, 
$+$ for D-branes with Dirichlet-Dirichlet boundary conditions for 
$X^{i'},\bar{X}^{i'}$ directions. In the Neumann-Dirichlet case one can
define the boundary state in the same way only if $\gamma$ is an integer or 
a half-integer. The above result (\ref{superfracten})
shows that a Dp-brane in the background has the ordinary tension $T_p$.

The structures of these D-branes are almost similar to those in the original 
Melvin 
background discussed in section 3. However we notice a 
remarkable property in this case:
{\sl under the condition that
the supersymmetry in the closed string theory is preserved 
($\beta_1=\pm\beta_2$ and $q_1=\pm q_2$) 
the open string one-loop amplitude (\ref{VAH}) vanishes.} Thus the 
D-branes in this special background are stable BPS objects. In the orbifold
limit\footnote{Equivalently one can take the limit 
$R\to 0,\ q_1R=\pm q_2 R=\f{k}{N},\ \beta_1=\beta_2=0$ by 
T-duality.} $R\to \infty$ with
\ba \f{\beta_1\al}{R}=\pm \f{\beta_2\al}{R}
=\f{k}{N},\ \ \ q_1=q_2=0, \label{LMH}
\ea 
these 
D-branes are identified with BPS fractional D-branes \cite{DoMo,DiDoGo} 
in the 
supersymmetric ALE 
orbifolds ${\bf C}^2/{\bf Z}_N$. 
More generally, if 
$\frac{\beta_i\al}{R}=\frac{k_i}{N}~(i=1,2)$ 
and if $k_1+k_2$ is even, the D-brane in the 
limits becomes a fractional
D-brane in type II (not necessarily supersymmetric) orbifolds 
${\bf C}^2/{\bf Z}_N$. On the other hand, if $k_1+k_2$ is odd, it is 
divided into 
an electric and a magnetic fractional D-brane \cite{BiCrRo} in 
type 0 orbifolds ${\bf C}^2/{\bf Z}_N$.

According to the parameter values $q_i R$ and 
$\frac{\beta_i\al}{R}$ $(i=1,2)$ the D-brane spectrum
dramatically changes again. 
For the irrational values only the D-branes which are stuck at the 
fixed point are allowed, while 
for the rational values there also exist the D-branes which can move around.
The latter are made of $N$ `fractional' D-branes, 
which can be seen as a generalization of a bulk D-brane in the orbifold theory
${\bf C}^2/{\bf Z}_N$ \cite{DoMo}. The detailed arguments of these D-branes 
are almost the same as before.

\subsection{Comments on D-brane Charges and K-theory}
\hspace{5mm}
Before we conclude this paper, let us briefly mention
the D-brane charges in the
supersymmetric Melvin background. It is known that 
D-brane charges in type IIB and IIA theory
are generally classified by K-theory $K^0(X)$ and $K^1(X)$ 
\cite{Witten3,Ho}, where $X$ 
represents the manifold of the spacetime considered. First we assume 
that the parameters are 
given by (\ref{LMH}). Then it is easy to see that the number of 
massless RR-fields are given by one for finite $R$ and by $N$ for the 
orbifold limit $R\to \infty$. In the orbifold language each of 
these $N$ RR-fields
comes from one untwisted sector and $N-1$ twisted sectors. Naively one may
think that the number of types of D$p$-branes should be given by $N$ 
in accordance with the $N$-types of boundary states (\ref{phase'}). 
However one should remember that for finite $R$ the type of a 
fractional D-brane changes if it goes around the compactified circle. 
This shows that the types of fractional D-branes (twisted RR-charge) 
do not preserve 
unless we take the limit $R\to \infty$. Thus we conclude that the D-brane
charge or equally the corresponding K-group\footnote{
The T-duality of the Melvin background is represented in terms of 
K-group as $K^0(M_5)=K^1(\hat{M}_5)$, where $\hat{M}_5$ is the T-dual of
the five dimensional Melvin background $M_5$.} should be given by the rather
trivial result
$K^0(M_5)=K^1(M_5)={\bf Z}$ for finite $R$.
On the other hand if we take the orbifold limit $R\to \infty$, then the
K-group should be given by the equivalent K-theory
$K^0_{{\bf Z}_N}({\bf R}^4)={\bf Z}^N$ \cite{Witten3,Ho}, 
which corresponds to the $N$-types of 
fractional D-branes on the orbifold ${\bf C}^2/{\bf Z}_N$. 
Thus the spectrum 
of D-brane charges does not appear to be continuously connected 
in the orbifold limit even 
though the masses of twisted RR-fields 
which couple D-branes continuously change. It would be 
interesting for the above facts to be understood from 
the viewpoint of the 
K-theory with $H$-flux like the twisted K-theory $K_{H}(X)$ discussed 
in \cite{FrSc,MaMoSe}.

Next we would like to consider the case where $\frac{\beta_i\al}{R}$ is 
irrational with $q_1=q_2=0$.
Then we find that in the limit $R\rightarrow\infty$ 
(a `large $N$ limit of the orbifold' \cite{TaUe})
there seems to be infinitely many massless RR-fields 
absorbing the zeromode along the circle ${\bf S}^1$. 
Thus we may have the `K-group' 
$K^0(M_5)=K^1(M_5)={\bf Z}^\infty$ in this limit even though for the
irrational parameters the space $M_5$ is no longer a smooth manifold
nor a `good' singular manifold like ordinary orbifolds.

It may not be so nonsense even to ask whether one can explain D-brane
charges in type 0 theory from the viewpoint of K-theory if one remembers 
that the type II Melvin background approaches type 0 theories 
with the appropriate limits.

\section{Conclusions and Discussions}
\setcounter{equation}{0}
\hspace{5mm}
In this paper we have investigated various aspects of D-branes 
in Melvin backgrounds by employing the boundary state formalism.
The boundary states are constructed in terms of the free field 
representations which can be obtained by applying T-duality to
the original sigma model of the Melvin backgrounds. Then we can
impose the ordinary Neumann or Dirichlet boundary conditions on the
free fields, though the resulting D-branes are geometrically non-trivial 
in the Melvin backgrounds.

The D-brane spectrum turns out to depend sensitively on whether the 
magnetic parameters $qR$ and $\f{\beta\al}{R}$ are rational or irrational.
In the former case the boundary states have the structure similar 
to orbifold theories and indeed they become the fractional D-branes\footnote{
More precisely, we have explicitly constructed D-branes in the 
Melvin background which are 
reduced to fractional D0-branes (D1-branes) in 
one of the orbifold limits $R\to \infty$ ($R\to 0$) in section 3. 
We left the discussion on the other limit as a future problem.} in 
the orbifolds ${\bf C}^n/{\bf Z}_N$ if we take the large radius (or
small radius in T-dual picture) limits.
For finite radius the $N$ kinds of `fractional D-branes' which belong 
to the different
irreducible representations of ${\bf Z}_N$ can be changed into each other by 
the monodromy if they go around the compactified circle. Even though
a single (`fractional') 
D-brane cannot move away from the fixed point (or the origin),
$N$ different kinds of `fractional' D-branes can move around as a unit 
(a `bulk' D-brane).
On the other hand, in the case of irrational parameters one needs the 
infinite number of the `fractional' D-branes 
in order to move away from the fixed point. 
We left the possibility of the existence 
of such D-branes as an unresolved problem.

The orbifold theory we obtain in the limits is defined in either 
type 0 or type II string theory depending on the rational parameters. 
In the type 0 case we found that a D-brane behaves like a combined
system of an electric fractional D-brane and a magnetic fractional D-brane. 

The individual effects of two parameters $q$ and $\beta$ are
roughly as follows. 
The non-zero value of the parameter $q$ twists the plane ${\bf R}^2$ 
spirally along the circle ${\bf S}^1$. As a result the world-volume of 
a D-brane which wraps the circle is bended along a spiral geodesic 
line (see eq.(\ref{metric2})). 
The effect of non-zero $\beta$, which means the non-zero $H$-flux,
is more complicated because one should mix the free fields
with its T-dualized ones as in eq.(\ref{newval1}) and (\ref{xx})
in order to return back to the original Melvin sigma model (\ref{SG1}).
This complexity may be due to the presence of $H$-flux.
Owing to this T-duality transformation, a bulk D0-brane at $\rho\neq 0$
in terms of
the free fields $(X',\bar{X}',Y')$ is interpreted as a 
D2-D0 bound state which wraps the two dimensional torus ($\rho=\mbox{fixed}$)
in the
original Melvin background (\ref{KK}). 
Interestingly, this two dimensional torus can be regarded as a 
non-commutative torus $\mathcal{A}_{\theta}$ with 
the noncommutativity $\theta=\f{\beta \al}{R}$ if we assume that the 
$B$-field is large $\beta\rho \gg 1$ and apply the argument 
in \cite{SeWi}. The observed dependence of D-branes on the parameters 
can be now reinterpreted as that on the parameter of the 
non-commutative torus $\mathcal{A}_{\theta}$. The corresponding 
operator algebra K-theory $K(\mathcal{A}_{\theta})={\bf Z}+\theta{\bf Z}$ 
\cite{PR}
shows the analogous difference between the rational and 
the irrational case, where the value ${\bf Z}+\theta{\bf Z}(\in {\bf R})$ 
means the
dimension of the corresponding projective module.
Indeed two ${\bf Z}$ charges
in the K-group
$K(\mathcal{A}_{\theta})$ represent the D2-brane and the D0-brane charges
as is clearly explained from the viewpoint of the tachyon condensation in
the open string theory \cite{NCT}. It would be interesting to 
examine the noncommutative
algebra on such D-branes by using the free field calculations since 
this background possesses non-zero $H$-flux (for a general discussion
on the relation between $H$-flux and noncommutative geometry see 
\cite{CoSc}).

Such an interpretation of a D0-brane in the free field model (\ref{SG3})
as a D2-D0 bound state in the Melvin sigma model (\ref{SG1}) seems to 
raise some intriguing questions. The first question is whether an ordinary 
D0-brane can also exist in the Melvin background (\ref{KK}) and what is
the form of its boundary state if it exists.
The second is why such a D2-D0 bound state can exist in the Melvin background
which does not wrap any homologically non-trivial cycle.
We leave these issues as future problems. 

It is also an interesting result that the boundary states of D-branes 
depend only on either of two parameters. 
This phenomenon leads to the strange Bose-Fermi degeneracy on specific 
D-branes even in nonsupersymmetric string backgrounds. We pointed out the 
interpretation of this as a remnant of the spontaneously broken supersymmetry.

Finally we would like to note that we can apply the above 
results for the original Melvin background (\ref{KK}) to its higher 
dimensional generalizations without any serious modifications
as we have seen in section 4. 
For specific values of parameters the model preserves some 
supersymmetries and the D-branes there become BPS states. Thus it would
be interesting to explore their supersymmetric world-volume theories
and their statuses from the viewpoint of the string dualities.

\begin{center}
\noindent{\large \bf Acknowledgments}
\end{center}
We are grateful to Y. Hikida, K. Ichikawa, H. Kajiura and E. Ogasa for
useful discussions.  
T.T. is supported by JSPS Research Fellowships for Young 
 Scientists.

\begin{center}
\noindent{\normalsize \bf Note Added}
\end{center}
After completing this paper we noticed the paper \cite{DuMo} on the net, 
which has some overlaps with our paper.
\appendix

\section{T-duality for Kaluza-Klein Background}\label{Ap:KK}
\setcounter{equation}{0}
\hspace{5mm}
The coordinate of world-sheet is $z=\sigma_1+i\sigma_2$ and we define 
its partial derivative as 
$\de=\f{1}{2}(\de_{1}-i\de_{2}), \db=\f{1}{2}(\de_{1}+i\de_{2})$.
Let us first consider the following bosonic sigma model:
\ba
S=\f{1}{\pi\al}\int d^2\sigma\left[(G_{ij}(X)+B_{ij}(X))
\de X^i \db X^j+\f{1}{4}\al R^{(2)}\phi(X)\right].
\ea
After substituting the Kaluza-Klein background like (\ref{KK}) we obtain
\ba
S&=&\f{1}{\pi\al}\int d^2\sigma\Bigl[
(G_{\mu\nu}(X)+B_{\mu\nu}(X))\db X^\nu\de X^\mu +e^{2\sigma(X)}
(\db Y+A_{\mu}\db X^\mu)(\de Y+A_{\mu}\de X^\mu) \no
&&\ \ \ \ \ \ \ \ \ \ \
+B_{\mu}(\db Y\de X^\mu-\db X^\mu\de Y)+\f{1}{4}\al R^{(2)}\phi(X)\Bigr].
\label{KKS}
\ea

Then we can perform T-duality along $Y$ direction (${\bf S^1}$ with
radius $R$) since the
fields $G_{\mu\nu}(X),B_{\mu\nu}(X),\phi(X)$ do not depend on $Y$ \cite{GPR}. 
Introducing the auxiliary vector field $V,\bar{V}$ we can rewrite 
(\ref{KKS}) as follows (we show only nontrivial parts)
\ba
\label{actionA}
S&=&\f{1}{\pi\al}\int d^2\sigma\Bigl[e^{2\sigma(X)}
(\bar{V}+A_{\mu}\db X^\mu)(V+A_{\mu}\de X^\mu)+
B_{\mu}(\bar{V}\de X^\mu -\db X^\mu V)\no
&&\ \ \ \ \ \ \ \ \ \ \ \ +(\bar{V}\de \ti{Y}-\db \ti{Y}V)
\Bigr],
\ea
where the new field $\ti{Y}$ is compactified on a circle 
with the radius $\f{\al}{R}$.
If we first integrate $\ti{Y}$, then we obtain 
$\db V-\de \bar{V}=0$ and the vector field $V, \bar{V}$ can be written as 
$V=\de Y,\ \bar{V}=\db Y$. Indeed one can easily see that 
this field $Y$ has
the periodicity $Y\sim Y+2\pi R$ as expected\footnote{To 
see this, one should note the relation
$d\ti{Y}=$(closed form)+(non-trivial cohomology). The non-trivial part is
discretized due to the periodicity $\ti{Y}\sim\ti{Y}+2\pi\f{\al}{R}$.
Then the term $\sim \int d\ti{Y}\wedge V$ in (\ref{actionA})
leads to the weight 
$\exp(i\f{\ti{w}}{R}\int_{C} V)$, where $\ti{w}$ is 
the winding number of the field
$\ti{Y}$ and $C$ is a one-cycle of the world-sheet. 
Since one should take a 
summation over $w$, the integration is 
quantized as $\int_{C} V\in 2\pi R{\bf Z}$. This shows that 
the period of $Y$ is $2\pi R$.}.
Next it is straightforward to integrate out $V$ first and one obtains
\ba
S&=&\f{1}{\pi\al}\int d^2\sigma\Bigl[\ddd+e^{-2\sigma(X)}
(\db \ti{Y}+B_{\mu}\db X^\mu)(\de \ti{Y}+B_{\mu}\de X^\mu) 
+A_{\mu}(\db \ti{Y}\de X^\mu-\db X^\mu\de \ti{Y})\no 
&&\ \ \ \ \ \ \ \ \ \ \ \ +\f{1}{4}\al R^{(2)}(\phi(X)-\sigma(X))\Bigr],
\label{KKS2}
\ea
where the shift of the dilaton field can be determined from the condition of
the conformal invariance (vanishing beta-function) \cite{GPR}. 
Thus we have obtained the following T-duality transformation:
\ba
\sigma'(X)=-\sigma(X),\ \ A_{\mu}'(X)=B_{\mu}(X),\ \ 
B_{\mu}'(X)=A_{\mu}(X),
\ \phi'(X)=\phi(X)-\sigma(X).
\ea

Note also that the equation of motion of (\ref{KKS}) is equivalent to 
that of 
(\ref{KKS2}) via the rule:
\ba
\de \ti{Y}&=&-B_\mu\de X^\mu-e^{2\sigma}(\de Y+A_{\mu}\de X^\mu), \no
\db \ti{Y}&=&-B_\mu\db X^\mu+e^{2\sigma}(\db Y+A_{\mu}\db X^\mu).
\label{rule}
\ea

In this paper we discuss superstring models and therefore we need
the supersymmetric
generalization of the above arguments. The simplest way to do this is 
to use the $N=1$ superspace formalism. One has only to replace $\de$ and $\db$ 
with the super covariant derivatives $D_{\theta}=\de_{\theta}+\theta\de$
 and $D_{\bar{\theta}}
=\de_{\bar{\theta}}+\bar{\theta}\db$ and 
replace the 
bosonic vector field $V$ with the fermionic vector super field $W$. The 
bosonic scalar field $X(z,\bar{z})$ should also be changed into
${\bf{X}}(z,\bar{z})
=X(z,\bar{z})+i\theta\psi'_{L}(z)+i\bar{\theta}\psi'_{R}(\bar{z})+\ddd$.
The
calculations are almost the same as before. For example, eq.(\ref{rule})
is replaced with
\ba
D_{\theta} \ti{Y}&=&-B_\mu D_{\theta} 
X^\mu-e^{2\sigma}(D_{\theta} Y+A_{\mu}D_{\theta} X^\mu), \no
D_{\bar{\theta}} \ti{Y}
&=&-B_\mu D_{\bar{\theta}} X^\mu+e^{2\sigma}(D_{\bar{\theta}}
 Y+A_{\mu}D_{\bar{\theta}} X^\mu).
\label{srule}
\ea

Let us see a useful example: T-duality of the background (\ref{KK}).
Utilizing the above arguments we can transform $q$ into $\beta$ by
T-duality. This fact is 
reconfirmed in the mass spectrum (\ref{closedmass}).

\section{Mode Expansion}\label{Ap:Mo}
\setcounter{equation}{0}
\hspace{5mm}

As we have seen in section 2, 
the theory can be represented by the free bosonic fields
$X',\bar{X}',\ Y'$ and their superpartners $\psi'_{L,R},\bar{\psi}'_{L,R},
\ \eta'_{L,R}$ with the boundary conditions 
(\ref{y'w}), (\ref{zero}) and (\ref{bc}). 
Thus we can find the following mode expansions
\ba
\label{modeex}
&&X'(z,\bar{z})=X'_{L}(z)+X'_{R}(\bar{z})
=i\s{\al}\sum_{m}\f{1}{m-\gamma}\ap_{m-\gamma}z^{-m+\gamma}
+i\s{\al}\sum_{m}
\f{1}{m+\gamma}\tilde{\ap}_{n+\gamma}\bar{z}^{-m-\gamma},\no
&&\bar{X}'(z,\bar{z})=\bar{X}'_{L}(z)+\bar{X}'_{R}(\bar{z})
=i\s{\al}\sum_{m}\f{1}{m+\gamma}\bar{\ap}_{m+\gamma}z^{-m-\gamma}
+i\s{\al}\sum_{m}
\f{1}{m-\gamma}\bar{\tilde{\ap}}_{m-\gamma}\bar{z}^{-m+\gamma},\no
&&\psi'_{L}(z)=\s{\al}\sum_{r}\psi_{r-\gamma}z^{-r+\gamma-\f12},\ \ \ 
\psi'_{R}(\bar{z})
=\s{\al}\sum_{r}\tilde{\psi}_{r+\gamma}\bar{z}^{-r-\gamma-\f12},\no
&&\bar{\psi}'_{L}(z)=
\s{\al}\sum_{r}\bar{\psi}_{r+\gamma}z^{-r-\gamma-\f12},\ \ \ 
\bar{\psi}'_{R}(\bar{z})
=\s{\al}\sum_{r}\bar{\tilde{\psi}}_{r-\gamma}\bar{z}^{-r+\gamma-\f12},\no
&&Y'(z,\bar{z})=y'-i\f{\al}{2}P'_L \ln z-i\f{\al}{2}P'_R \ln \bar{z}
+i\s{\f{\al}{2}}\sum_{m\neq 0}\f{1}{m}\beta_{m}z^{-m}
+i\s{\f{\al}{2}}\sum_{m\neq 0}\f{1}{m}\tilde{\beta}_{m}\bar{z}^{-m},\no
&&\eta'_{L}(z)=\s{\al}\sum_{r}\eta_{r}z^{-r-\f12},\ \ \ 
\eta'_{R}(\bar{z})=\s{\al}\sum_{r}\ti{\eta}_{r}\bar{z}^{-r-\f12},
\ea
and (anti)commutation rules
\ba
&&[\ap_{m-\gamma},\bar{\ap}_{n+\gamma}]=(m-\gamma)\delta_{m,-n},\ \ \ 
[\ti{\ap}_{m+\gamma},\ti{\bar{\ap}}_{n-\gamma}]=(m+\gamma)\delta_{m,-n},\no
&&\{\psi_{r-\gamma},\bar{\psi}_{s+\gamma}\}=\delta_{r,-s},
\ \ \ \{\ti{\psi}_{r+\gamma},\ti{\bar{\psi}}_{s-\gamma}\}
=\delta_{r,-s},\no
&&[\beta_{m},\beta_{n}]=m\delta_{m,-n},\ \ \
[\ti{\beta}_{m},\ti{\beta}_{n}]=m\delta_{m,-n},\no
&&\{\eta_{r},\eta_{s}\}=\delta_{r,-s},\ \ \ 
\{\ti{\eta}_{r},\ti{\eta}_{s}\}=\delta_{r,-s}.
\ea

The ${\cal N}=1$ super-Virasoro generators 
$L_m, \tilde{L}_m, G_r, \tilde{G}_r$ are obtained in the usual way 
because the action is written by free fields
\ba
\label{eq5.5}
L_m&=&:\frac{1}{2}\sum_k\beta_{m-k}\beta_k+\sum_k
\bar{\alpha}_{m-k+\gamma}\alpha_{k-\gamma}\no
&&+\frac{1}{2}\sum_r(r-\frac{m}{2})\eta_{m-r}\eta_r
+\sum_r(r-\gamma-\frac{m}{2})\bar{\psi}_{m-r+\gamma}\psi_{r-\gamma}:~,\no
G_r&=&\sum_k(\beta_k\eta_{r-k}+\alpha_{k-\gamma}\bar{\psi}_{r-k+\gamma}
+\bar{\alpha}_{k+\gamma}\psi_{r-k+\gamma}),
\ea
where $:\sim:$ is the conformal normal ordering. For example 
for $0<\gamma<\frac{1}{2}$, $L_0$ for NS-sector becomes\footnote{Note 
that if $\gamma$ is out of this region, 
the above $L_0$ is not positive definite, therefore we have to redefine the 
ground state correctly.} 
\ba
\label{eq6}
L_0&=&\frac{\al}{4}P'^2_L+\sum_{k=1}^{\infty}\beta_{-k}\beta_k
+\sum_{k=1}^{\infty}\bar{\alpha}_{-k+\gamma}\alpha_{k-\gamma}
+\sum_{k=0}^{\infty}\alpha_{-k-\gamma}\bar{\alpha}_{k+\gamma}\no
&&+\sum_{r=1/2}^{\infty}r\eta_{-r}\eta_r
+\sum_{r=1/2}^{\infty}(r-\gamma)\bar{\psi}_{-r+\gamma}\psi_{r-\gamma}
+\sum_{r=1/2}^{\infty}(r+\gamma)\psi_{-r-\gamma}\bar{\psi}_{r+\gamma}
+\frac{\gamma}{2},
\ea
where $P'_L$ is given by (\ref{zero}).
Note that here we abbreviate the contributions from other directions 
than $Y', X'$ and $\bar{X}'$. $L_0$ for R-sector is almost the same
except that 
$r$ runs integer values and that the zero point energy shifts. 
The antiholomorphic 
components $\tilde{L}_m, \tilde{G}_r$ is written in the same form.

For the later use, 
we define the operators $\hat{N}_{L},\hat{N}_{R}$ (we show only 
the result in NSNS sector)
\ba
\label{NLR}
\hat{N}_{L}
&=&\sum_{k=1}^\infty\f{k}{k-\gamma}\bar{\ap}_{-k+\gamma}\ap_{k-\gamma}
+\sum_{k=1}^{\infty}\f{k}{k+\gamma}\ap_{-k-\gamma}\bar{\ap}_{k+\gamma}
+\sum_{k=1}^\infty\beta_{-k}\beta_{k}\no
&&+\sum_{r=\f12}^{\infty}r\bar{\psi}_{-r+\gamma}\psi_{r-\gamma}
+\sum_{r=\f12}^{\infty}r\psi_{-r-\gamma}\bar{\psi}_{r+\gamma}-\frac{1}{2},\no
\hat{N}_{R}
&=&\sum_{k=1}^\infty\f{k}{k-\gamma}\ti{\ap}_{-k+\gamma}\bar{\ti{\ap}}
_{k-\gamma}+\sum_{k=1}^{\infty}\f{k}{k+\gamma}\bar{\ti{\ap}}_{-k-\gamma}
\ap_{k+\gamma}+\sum_{k=1}^\infty\ti{\beta}_{-k}\ti{\beta}_{k}\no
&&+\sum_{r=\f12}^{\infty}r\ti{\psi}_{-r+\gamma}\bar{\ti{\psi}}_{r-\gamma}
+\sum_{r=\f12}^{\infty}r\bar{\ti{\psi}}_{-r-\gamma}
\ti\psi_{r+\gamma}-\frac{1}{2}.
\ea 
The last constant term $-1/2$ is replaced by $0$ for RR-sector. 

The angular momentum operators $\hat{J}_{L},\hat{J}_{R}$, 
are also defined to be (we show the result for NSNS-sector with $0<\gamma<1/2$)
\ba
\label{JLR}
\hat{J}_{L}\!\!\!&=&\!\!\!-\sum_{k=1}^{\infty}\f{1}{k-\gamma}
\bar{\ap}_{-k+\gamma}\ap_{k-\gamma}+\!\!
\sum_{k=0}^{\infty}\f{1}{k+\gamma}\ap_{-k-\gamma}\bar{\ap}_{k+\gamma}
-\!\!\sum_{r=\f12}^{\infty}\bar{\psi}_{-r+\gamma}\psi_{r-\gamma}
+\!\!\sum_{r=\f12}^{\infty}\psi_{-r-\gamma}\bar{\psi}_{r+\gamma}+\f12,\no
\hat{J}_{R}\!\!\!&=&\!\!\!\sum_{k=1}^\infty\f{1}{k-\gamma}\ti{\ap}_{-k+\gamma}
\bar{\ti{\ap}}_{k-\gamma}
-\!\!\sum_{k=0}^{\infty}\frac{1}{k+\gamma}\bar{\ti{\ap}}_{-k-\gamma}
\ap_{k+\gamma}
+\!\!\sum_{r=\f12}^{\infty}\ti{\psi}_{-r+\gamma}\bar{\ti{\psi}}_{r-\gamma}
-\!\!\sum_{r=\f12}^{\infty}\bar{\ti{\psi}}_{-r-\gamma}
\ti\psi_{r+\gamma}-\f12.\no
\ea
The last constant term for RR-sector is the same as the above.

\section{Formulae of $\theta$-functions}\label{Ap:TF}
\setcounter{equation}{0}
\hspace{5mm}
Here we summarize the formulae of $\theta$-functions. First define the 
following $\theta$-functions:
\ba
\eta(\tau)&=&q^{\f{1}{24}}\prod_{n=1}^{\infty}(1-q^n),\no
\theta_{1}(\nu|\tau)&=&2q^{\f18}\sin(\pi\nu)\prod_{n=1}^{\infty}(1-q^n)
(1-e^{2i\pi\nu}q^{n})(1-e^{-2i\pi\nu}q^{n}),\no
\theta_{2}(\nu|\tau)&=&2q^{\f18}\cos(\pi\nu)\prod_{n=1}^{\infty}(1-q^n)
(1+e^{2i\pi\nu}q^{n})(1+e^{-2i\pi\nu}q^{n}),\no
\theta_{3}(\nu|\tau)&=&\prod_{n=1}^{\infty}(1-q^n)
(1+e^{2i\pi\nu}q^{n-\f12})(1+e^{-2i\pi\nu}q^{n-\f12}),\no
\theta_{4}(\nu|\tau)&=&\prod_{n=1}^{\infty}(1-q^n)
(1-e^{2i\pi\nu}q^{n-\f12})(1-e^{-2i\pi\nu}q^{n-\f12}), \label{th}
\ea
where we have defined $q=e^{2i\pi\tau}$.

Next we show the modular properties as follows
\ba
\eta(\tau)&=&(-i\tau)^{-\f12}\eta(-\f{1}{\tau}),\ \ \theta_{1}(\nu|\tau)
=i(-i\tau)^{-\f12}e^{-\pi i\f{\nu^2}{\tau}}\theta_{1}
(\nu/\tau|-\f{1}{\tau}), \no
\theta_{2}(\nu|\tau)&=&(-i\tau)^{-\f12}e^{-\pi i\f{\nu^2}{\tau}}
\theta_{4}(\nu/\tau|-\f{1}{\tau}), \ \ \theta_{3}(\nu|\tau)
=(-i\tau)^{-\f12}e^{-\pi i\f{\nu^2}{\tau}}\theta_{3}(\nu/\tau|-\f{1}{\tau})
, \no\theta_{4}(\nu|\tau)&=&(-i\tau)^{-\f12}e^{-\pi i\f{\nu^2}{\tau}}
\theta_{2}(\nu/\tau|-\f{1}{\tau}). \label{TF}
\ea

Their quasi periodicities are also given by
\ba
\label{period}
\theta_1(\nu+\tau|\tau)&=&-e^{-2\pi i\nu-\pi i \tau}\theta_1(\nu|\tau), \no
\theta_2(\nu+\tau|\tau)&=&e^{-2\pi i\nu-\pi i \tau}\theta_2(\nu|\tau), \no
\theta_3(\nu+\tau|\tau)&=&e^{-2\pi i\nu-\pi i \tau}\theta_3(\nu|\tau), \no
\theta_4(\nu+\tau|\tau)&=&-e^{-2\pi i\nu-\pi i \tau}\theta_4(\nu|\tau). 
\ea

It is useful to note the relation
\ba
\prod_{a=1}^{4}\theta_{3}(\nu_a|\tau)-\prod_{a=1}^{4}\theta_{2}(\nu_a|\tau)
-\prod_{a=1}^{4}\theta_{4}(\nu_a|\tau)+\prod_{a=1}^{4}\theta_{1}(\nu_a|\tau)
=2\prod_{a=1}^{4}\theta_{1}(\nu'_a|\tau), \label{TF2}
\ea
where we have defined
\ba
&&2\nu'_1=\nu_1+\nu_2+\nu_3+\nu_4,\ \ 2\nu'_2=
\nu_1+\nu_2-\nu_3-\nu_4,\no \ \ 
&&2\nu'_3=\nu_1-\nu_2+\nu_3-\nu_4, \ \ 2\nu'_4=\nu_1-\nu_2-\nu_3+\nu_4.
\ea

\end{document}